\documentstyle[12pt,amsmath,amssymb,epsf,epic,cite]{article}
\numberwithin{equation}{section}

\newcounter{resultcounter}[section]
\renewcommand{\theresultcounter}{\arabic{section}.\arabic{resultcounter}}
\newtheorem{thm}[resultcounter]{Theorem}

\newtheorem{prop}[resultcounter]{Proposition}
\newtheorem{cor}[resultcounter]{Corollary}

\newcommand{\fer}[1]{(\ref{#1})}
\newcommand{\scalprod}[2]{\left\langle {#1}, {#2}\right\rangle}
\newcommand{\bbbone}{\mathchoice {\rm 1\mskip-4mu l} {\rm 1\mskip-4mu l}
{\rm 1\mskip-4.5mu l} {\rm 1\mskip-5mu l}}

\newcommand{\s}{{\rm S}}
\renewcommand{\r}{{\rm R}}
\newcommand{\R}{{\mathbb R}}
\newcommand{\cx}{{\mathbb C}}
\newcommand{\rx}{{\mathbb R}}
\renewcommand{\d}{{\rm d}}
\newcommand{\e}{{\rm e}}
\newcommand{\fm}{{\frak M}}

\renewcommand{\i}{{\rm i}}
\newcommand{\dom}{{\cal D}}
\newcommand{\qed}{{\hfill $\square$}}
\newcommand{\cD}{{\cal D}}
\newcommand{\cH}{{\cal H}}

\begin{document}

\title{Overlapping Resonances in\\
 Open Quantum Systems}
\author{{}\\
Marco Merkli\footnote{Department of Mathematics and Statistics, Memorial University of Newfoundland, St. John's, NL, Canada, A1C 5S7}\ \,\footnote{Email: merkli@mun.ca, webpage: http://www.math.mun.ca/$\sim$merkli/, supported by an NSERC Discovery Grant} \qquad Haifeng Song$^*$\footnote{Email: hs1858@mun.ca}
}
\maketitle

\begin{abstract}
An $N$-level quantum system is coupled to a bosonic heat reservoir at positive temperature. We analyze the system-reservoir dynamics in the following regime:  The strength  $\lambda$ of the system-reservoir coupling is fixed and small, but larger than the spacing $\sigma$ of system energy levels. For vanishing $\sigma$ there is a manifold of invariant system-reservoir states and for $\sigma>0$ the only invariant state is the joint equilibrium. The manifold is invariant for $\sigma=0$ but becomes quasi-invariant for $\sigma>0$. Namely, on a first time-scale of the order $1/\lambda^2$, initial states approach the manifold. Then they converge to the joint equilibrium state on a much larger time-scale of the order $\lambda^2/\sigma^2$. We give a detailed expansion of the system-reservoir evolution showing the above scenario.
\end{abstract}

\section{Introduction and main results}

We consider an open quantum system consisting of a small, finite-dimensional part interacting with a heat bath, modeled by a spatially infinitely extended free Bose gas in thermal equilibrium. The analysis of such systems, and especially of their dynamics, has a long tradition. The reduced dynamics of the small system alone is described in the theoretical physics literature primarily using master equation techniques, which rely on approximations that are not controlled mathematically, but are very popular and successful nevertheless  \cite{JZKGKS,Schl,GZ, BreuerPetruccione,NC}. A rigorous approach is the van Hove, or weak coupling limit \cite{Davies1,Davies2,AL}. It describes the dynamics of the small system for times up to the order of $\lambda^{-2}$, where $\lambda$ is the strength of the system-environment coupling. Given a fixed $\lambda$, the time-asymptotics, $t\rightarrow\infty$, cannot be resolved with the weak coupling method. It is shown in \cite{DK} however that, for a class of open systems, if the conditions for the weak coupling limit are satisfied, then the small subsystem converges to a final state in the long time limit.

The analysis of the {\em total} system -- the small system plus the reservoir -- is more delicate than that of the small subsystem alone. Over the last decade and a half, a perturbation theory based on quantum resonance methods has been developed to deal with this problem, see \cite{JP,BFS,M,JP1,DJ,FM,MMS,MMS1,MSB}. It is implemented in various forms, using spectral deformation, positive commutator and renormalization group techniques and permits a mathematically rigorous treatment of the full dynamics (system plus reservoir), for {\em fixed,  small} coupling $\lambda$ and for {\em all} times $t\geq 0$. Other than the spectral approach of the above references, the polymer expansion method of \cite{DK} allows the analysis the total system as well, see \cite{DK2}.

\medskip

The techniques of the above works are based on a perturbation theory in the system-reservoir coupling parameter $\lambda$. The latter is assumed to be small relative to the spacing $\sigma>0$ between the energy levels of the small system: $|\lambda|<\!\!<\sigma$. This is the {\em isolated resonances regime}. However, there are many physical systems for which this condition is not valid. For instance in {\em complex open systems}, the small system itself is composed of many  individual parts (particles) and the energy level spacing may become very small. Take the Hamiltonian of a system of $N$ spins, having $2^N$ eigenvalues. The total energy of the spins is of the order of $N$. The generic energy spacing is thus of the order of $\sigma\sim N/2^N$, which is exponentially small in $N$. For such systems, the condition  $|\lambda|<\!\!<\sigma$ is not reasonable. 

In the present work, we develop the resonance method in the {\em overlapping resonances regime} $\sigma<\!\!< |\lambda|$. We study here the simplest case, in which all the system energies lie close together relative to $|\lambda|$. Our results hold for a fixed, finite (but arbitrary) dimension $N$ of the small system and for small coupling constants,  $|\lambda|\leq\lambda_0$, for some $\lambda_0>0$.

The $N$-level system coupled to a thermal reservoir is described by the Hamiltonian 
$$
H^\Lambda(\sigma,\lambda) = \sigma H_\s +H_\r^\Lambda +\lambda G\otimes\Phi^\Lambda(g),
$$
acting on the Hilbert space ${\cx}^N \otimes{\cal F}(L^2(\Lambda,\d^3x))$, where the second factor is the Fock space over the one-particle Hilbert space of wave functions localized in a finite box $\Lambda\subset{\mathbb R}^3$. The system Hamiltonian $H_\s$ is an arbitrary self-adjoint operator on ${\cx}^N$. The reservoir Hamiltonian $H^\Lambda_\r$ is the second quantization of the single Boson energy, the self-adjoint Laplace operator with periodic boundary conditions. The system-reservoir interaction is the product of a self-adjoint $G$ acting on the system and the field operator $\Phi^\Lambda(g)=\frac{1}{\sqrt 2}(a^*(g)+a(g))$, where $a^*, a$ are the creation and annihilation operators on ${\cal F}(L^2(\Lambda,\d^3x))$, smoothed out with the form factor $g$ supported in $\Lambda$. The Hamiltonian contains the two parameters $\sigma\geq 0$ and $\lambda\in{\mathbb R}$, the system energy level splitting parameter and the interaction strength, respectively. The bosonic field is initially in its {\em thermal equilibrium} state at positive temperature $1/\beta$, given by the density matrix $\rho^\Lambda_{\r,\beta}\propto\e^{-\beta H_\r^\Lambda}$. In order to have a true open system, one performs the infinite-volume limit of the reservoir, in which the box $\Lambda$ grows to all of ${\mathbb R}^3$. More precisely, the expectation values of observables (Weyl operators) of the reservoir, in the thermal state, have a limit as $\Lambda\rightarrow{\mathbb R}^3$. This defines the infinite-volume equilibrium state $\omega_{\r,\beta}$ by its expectation values $\omega_{\r,\beta}(W(f))$ on the Weyl operators. A Hilbert space on which that state is represented by a vector can then be reconstructed using the Gelfand-Naimark-Segal (GNS) construction \cite{AW}. This procedure leads to the description of the coupled system as a $W^*$-dynamical system \cite{BR,OQS1}. It consists of a Hilbert space
\begin{equation}
\cH = \cH_\s\otimes\cH_\r,
\label{m2}
\end{equation}
of a von Neumann algebra of observables 
\begin{equation}
\fm = \fm_\s\otimes \fm_\r,
\label{m3}
\end{equation}
and of a Heisenberg dynamics of $\fm$,
\begin{equation}
A\mapsto \alpha_{\sigma,\lambda}^t(A) = \e^{\i tL(\sigma,\lambda)} A\e^{-\i tL(\sigma,\lambda)}, \qquad A\in\fm.
\label{fullLiouville}
\end{equation}
The Liouvillian $L(\sigma,\lambda)$ is a self-adjoint operator on $\cH$. 
The small system is an $N$-level system having a Hamiltonian $H_\s$. In the GNS (Gelfand-Naimark-Segal) representation, the Hilbert space is $\cH_\s=\cx^N\otimes\cx^N$ and the algebra of observables is given by $\fm_\s = {\cal B}({ \cx^N})\otimes\bbbone_{\cx^N}$ (bounded linear operators). The dynamics is implemented as
\begin{equation}
A_\s\mapsto \e^{\i t L_\s} (A_\s\otimes\bbbone_{\cx^N}) \e^{-\i tL_\s},\qquad A_\s\in {\cal B}(\cx^N),
\label{m10}
\end{equation}
where 
\begin{equation}
L_\s = H_\s\otimes\bbbone_{{\mathbb C}^N} - \bbbone_{{\mathbb C}^N}\otimes {\cal C}H_\s{\cal C}
\label{m11}
\end{equation}
is the self-adjoint system Liouville operator. Here, $\cal C$ is the operator taking the complex conjugate of components of vectors represented in  the orthonormal eigenbasis $\{\varphi_a\}_{a=1}^N$ of the interaction operator,
\begin{equation}
G\varphi_a=g_a\varphi_a,\quad a=1,\ldots,N.
\label{m22}
\end{equation}
The procedure of doubling of the Hilbert space is well known in the physics literature, also called the `Liouville Representation', see e.g. \cite[Chapter 3]{Muk}.

The reservoir state is the thermodynamic (infinite volume) limit of a free Bose gas in equilibrium at inverse temperature $\beta$. Its Hilbert space representation has first been constructed in \cite{AW} and a unitarily equivalent representation, suitable for the use of spectral translation techniques, has been given in \cite{JP}. The GNS Hilbert space is $\cH_\r = {\cal F}\big( L^2({\mathbb R}\times S^2,\d u\times\d\vartheta)\big) = \oplus_{n\geq 0}  L_{{\rm symm}}^2(({\mathbb R}\times S^2)^n,(\d u\times\d\vartheta)^n)$, 
the symmetric Fock space over the one-particle function space $L^2({\mathbb R}\times S^2,\d u\times\d\vartheta)$. Here, $\d\vartheta$ is the uniform measure on $S^2$. The thermal field operator is given by
\begin{equation}
\Phi(f_\beta)=\frac{1}{\sqrt 2}\big( a^*(f_\beta) +a(f_\beta)\big),
\label{m5}
\end{equation}
where $a^*(f_\beta)= \int_{{\mathbb R}\times S^2} f_\beta(u,\vartheta) a^*(u,\vartheta)\ \d u\d\vartheta$ is the creation operator acting on the Fock space $\cH_\r$ and $a(f_\beta)$ is its adjoint, smoothed out with $f_\beta\in L^2({\mathbb R}\times S^2,\d u\times\d\vartheta)$ defined by
\begin{equation}
f_\beta(u,\vartheta):=\sqrt{\frac{u}{1-\e^{-\beta u}}}\ |u|^{1/2}\left\{
  \begin{array}{ll}
   f(u,\vartheta), & \hbox{if}\,\, u\geq0, \\
    \overline{f}(-u,\vartheta), & \hbox{if}\,\, u<0.
  \end{array}
\right.
\label{glue}
\end{equation}
Here, $f\in L^2({\mathbb R}^3,\d^3k)$ is represented in polar coordinates (and in Fourier space). The thermal Weyl CCR algebra $\fm_\r\subset{\cal B}(\cH_\r)$ is the von Neumann algebra generated by the unitary Weyl operators $W(f_\beta):=\e^{\i\Phi(f_\beta)}$. The dynamics on $\fm_\r$ is given by the Bogoliubov transformation $t\mapsto W(\e^{\i t u}f_\beta) = \e^{\i t L_\r} W(f_\beta) \e^{-\i tL_\r}$. 
It is implemented by the self-adjoint reservoir Liouvillian
\begin{equation}
L_\r = \d\Gamma(u):= \int_{{\mathbb R}\times S^2} u\ a^*(u,\vartheta) a(u,\vartheta) \d u\d\vartheta,
\label{m8}
\end{equation}
the second quantization of the operator of multiplication by $u\in\mathbb R$. 
 The vacuum vector $\Omega_\r\in\cH_\r$ represents the $\beta$-KMS state w.r.t. the dynamics generated by \fer{m8}.

The Liouville operator $L(\sigma,\lambda)$ determining the full dynamics, \fer{fullLiouville}, has the form
\begin{equation}
L(\sigma,\lambda)=L_0(\sigma)+\lambda V,
\label{m11'}
\end{equation}
with a free part 
\begin{equation}
L_0(\sigma) = \sigma L_\s+L_\r
\label{m12}
\end{equation}
(see  \fer{m11}, \fer{m8}) and where the system-reservoir interaction is
\begin{equation}
\lambda V=\lambda G\otimes\bbbone_{{\mathbb C}^N}\otimes\Phi(g_\beta).
\label{m13}
\end{equation}
Here, $\sigma$ and $\lambda$ are two real parameters, $G$ is a self-adjoint matrix on ${\mathbb C}^N$ and $g_\beta\in L^2({\mathbb R}\times S^2)$ is obtained from a {\it form factor} $g\in L^2({\mathbb R}^3)$ using the relation \fer{glue}. It is well known that $L(\sigma,\lambda)$ is self-adjoint for all $\lambda,\sigma\in\mathbb R$  (this can be proven by the Glimm-Jaffe-Nelson commutator theorem, see e.g. \cite[Theorem A.2]{M}). We assume the following regularity of the form factor.
\medskip

{\bf Assumption A1.} (Analyticity) There is a $\theta_0>0$ such that $\theta\mapsto
g_\beta(u+\theta,\vartheta)$ has an analytic extension to the domain $\{\theta\in{\mathbb C}\ :\ |\theta|<\theta_0\}$, as a map from
$\mathbb C$ to $L^2({{\mathbb R}}\times S^2,\d u\times \d\vartheta)$.

\medskip

{\bf Assumption A2.} (Ultra-violet decay) There is an $\epsilon > 0$ such that $\e^{a|k|} g(k)\in L^2({\mathbb R}^3,\d^3k)$ for an $a>(1/2+\epsilon)\beta$, where $\beta$ is the inverse temperature.

\medskip
Examples of form factors satisfying this condition are $g(r,\vartheta)=r^p\e^{-a r^m}g_1(\vartheta)$ (polar coordinates in $\rx^3$), where $p=-1/2+n$, $n=1,2,\ldots$, $m=1,2$, and $g_1(\vartheta)\in\rx$ (see also \cite{FM} for more general classes of admissible $g$). More generally, we charaterize the infrared behaviour of the form factor by $p\geq -\frac{1}{2}$ satisfying $0<\lim_{|k|\rightarrow 0}\frac{|g(k)|}{|k|^p}=C<\infty$. The value of $p$ depends on the physical model considered. For quantum optical systems, $p=1/2$, for the quantized electromagnetic field, $p=-1/2$. We define the complex numbers 
\begin{equation}
\delta_{a,b}=-\textstyle\frac{1}{2}(g_a^2-g_b^2)\scalprod{g}{|k|^{-1}g}+\i \frac{\pi}{2} (g_a-g_b)^2
\left
\{\begin{array}{lll} 
0 & \mbox{\rm if $p>-1/2$}\\
 \xi(0)>0 & \mbox{\rm if $p=-1/2$}
\end{array},
\right.
\label{delta}
\end{equation}
for $a,b=1,\ldots,N$ and where 
\begin{equation}
\xi (0)=\lim_{\epsilon\downarrow
0}\frac{1}{\pi}\int_{{\mathbb{R}}^3}\coth(\frac{\beta|k|}{2})|g(k)|^2\frac{\epsilon}{|k|^2+\epsilon^2}\d^3k.
\label{xinot}
 \end{equation}
The $\lambda^2\delta_{a,b}$  are the {\em resonance energies} for $\sigma=0$, see Theorem \ref{thm3} below. The following assumption simplifies the presentation of our results. 
\medskip

 {\bf Assumption A3.} (Non-degeneracy) The spectrum $\{g_a\}_{a=1}^N$ of $G$ is such that all non-zero $\delta_{a,b}$ are distinct.
\medskip

Our analysis is readily generalized to the case of degenerate resonances (see the proof of Theorem \ref{thm4}). Indeed, we do this for the spin-boson model, in which the two non-zero resonances are given by $\delta_{1,2}=\delta_{2,1}=\i\frac\pi2 \xi(0)$.

The following is a well-coupledness condition which we will assume for some results. It implies that the coupled system has a unique stationary state (the coupled equilibrium).

\medskip

{\bf Assumption A4.} (Fermi Golden Rule Condition) For all $a,b$, $a\neq b$, we have ${\rm Im}\delta_{a,b}>0$ and $\scalprod{\varphi_a}{H_\s\varphi_b}\neq 0$.

\medskip

We show in Appendix A that the manifold of normal $\alpha_{0,\lambda}^t$-invariant states on $\frak M$ is the convex span of the states $\omega_{a}=\omega_{\s,a}\otimes\omega_{\r,a}$, $a=1,\ldots,N$. Here, $\omega_{\s,a}$ is given by the rank-one density matrix $|\varphi_a\rangle\langle\varphi_a|$ (spectral projection associated to $G$), and $\omega_{\r,a}$ is a normal perturbation of the reservoir equilibrium state, explicitly given in \fer{mexpl}. When $\sigma>0$ is small, then there is a unique (normal) $\alpha_{\sigma,\lambda}^t$-invariant state on $\frak M$, namely, the coupled system-reservoir equilibrium state $\omega_{\beta,\sigma,\lambda}$, which is an $(\alpha^t_{\sigma,\lambda},\beta)$-KMS state.

\medskip

Our main result, summarized in Theorem \ref{dynthm} below, concerns the dynamics of initial conditions and observables taken from sets ${\cal S}_0$ and ${\frak M}_0$, respectively. ${\cal S}_0$ is a set of bounded linear functionals on $\fm$ (defined in \fer{anstates}), dense in the set of all states of $\fm$. All states of the form $\omega_\s\otimes\omega_{\r,\beta}$ are in ${\cal S}_0$, where $\omega_\s$ is an arbitrary state on $\fm_\s$ and $\omega_{\r,\beta}$ is the equilibrium state of the reservoir. $\fm_0$ is the collection of translation analytic elements of $\fm$, a dense set in $\fm$, see \fer{anstates}.  All observables $A=A_\s\otimes\bbbone_\r$ of the system alone are in $\fm_0$. To express the details of the evolution, we introduce the following. For $a,b=1,\ldots,N$, $a\neq b$, set
\begin{eqnarray}
\lefteqn{
\eta_{a,b}(\sigma,\lambda) = \lambda^2\delta_{a,b} + \sigma\big([H_\s]_{a,a}-[H_\s]_{b,b}\big)} \nonumber\\
&& \qquad\qquad- \frac{\sigma^2}{\lambda^2}\Big(\sum_{c=1,\ldots,N; c\neq a}\frac{|[H_\s]_{a,c}|^2}{\delta_{c,b}-\delta_{a,b}}+ \sum_{c=1,\ldots,N; c\neq b}\frac{|[H_\s]_{b,c}|^2}{\delta_{a,c}-\delta_{a,b}}\Big),
\label{pertevnon-zero}
\end{eqnarray}
where $[H_\s]_{b,c}=\scalprod{\varphi_b}{H_\s\,\varphi_c}$ are the matrix elements of the system Hamiltonian. For $a=1,\ldots,N$, set
\begin{equation}
\eta_{a,a}(\sigma,\lambda) = 2\i\frac{\sigma^2}{\lambda^2}\xi_a,
\label{pertevzero}
\end{equation}
where $\xi_a\geq 0$ are the eigenvalues of the real symmetric $N\times N$ matrix $T$ with matrix elements
\begin{equation}
{}[T]_{a,b}=
\left\{
\begin{array}{cl}
\displaystyle -\frac{{\rm Im}\delta_{a,b}}{|\delta_{a,b}|^2} |[H_\s]_{a,b}|^2,  & \mbox{if $a\neq b$}\\
 & \\
\displaystyle \sum_{c=1,\ldots,N; c\neq a} \frac{{\rm Im}\delta_{a,c}}{|\delta_{a,c}|^2} |[H_\s]_{a,c}|^2, & \mbox{if $a=b$}.
\end{array}
\right.
\label{mmm6}
\end{equation}
The vector $\frac{1}{\sqrt N}(1,\ldots,1)$ is in the null space of $T$. We enumerate the eigenvalues of $T$ s.t. $\xi_1=0$. Under Assumption A4, zero is a simple eigenvalue of $T$ (see after \fer{mmm5} for a proof). We show in Theorem \ref{thm4} that, for $\sigma<\!\!<|\lambda|$, the resonances are given by
\begin{eqnarray}
\varepsilon_{a,b}( \sigma,\lambda)&=& \eta_{a,b}(\sigma,\lambda) + O\left(\sigma^2 |\lambda|^{-1}\right)+ O_\lambda(\sigma^3) \label{l7}\\
\varepsilon_a(\sigma,\lambda)&=& 2\i\frac{\sigma^2}{\lambda^2} \xi_a + O\left(\sigma^2 |\lambda|^{-1}\right)+ O_\lambda(\sigma^3).\label{17'}
\end{eqnarray}
Here, $O_\lambda(\sigma^3)$ is a term $f(\lambda,\sigma)$ satisfying $\limsup_{\sigma\rightarrow 0}\sigma^{-3}\|f(\lambda,\sigma)\|=C_\lambda<\infty$.

\begin{thm}[Dynamics in the overlapping resonances regime.]
\label{dynthm}
Assume A1-A4. There is a constant $\lambda_0>0$, such that for $0<|\lambda|<\lambda_0$, the following holds. There is a $\sigma_0>0$ (depending on $\lambda$) such that for $0\leq \sigma< \sigma_0$ and for any $\omega_0\in{\cal S}_0$, $A\in\fm_0$, $t\geq 0$, we have 
\begin{equation}
\omega_0\big(\alpha^t_{\sigma,\lambda}(A)\big) = \omega_{\beta,\sigma,\lambda}(A)+\sum_{a=2}^N \e^{\i t\varepsilon_a(\sigma,\lambda)} \chi_a(A) + \sum_{a,b=1\atop a\neq b}^N \e^{\i t\varepsilon_{a,b}(\sigma,\lambda)} \chi_{a,b}(A) +O(\e^{-\gamma t}).
\label{1}
\end{equation}
The $\chi_a$, $\chi_{a,b}$ in \fer{1} are linear functionals on $\fm_0$. They depend on $\sigma,\lambda$ and the initial condition $\omega_0$, but not on $t$. The decay rate $\gamma>0$ is independent of $\lambda,\sigma$ and satisfies $\gamma>\max\{{\rm Im}\varepsilon_a,{\rm Im}\varepsilon_{a,b}\}$.
\end{thm}

{\bf Discussion.\ } The imaginary parts ${\rm Im}\varepsilon_{a,b}\propto \lambda^2$ and ${\rm Im}\varepsilon_a\propto \sigma^2/\lambda^2$ (to leading order) have the associated decay times $t_1\propto \lambda^{-2}<\!\!< t_2\propto\lambda^2/\sigma^2$. The representation \fer{1} thus paints the following picture. In the non-degenerate situation, $\sigma>0$, the remainder term becomes negligible very quickly, for $t> t_0=1/\gamma$. Then, for $t> t_1$ the sum over the $\chi_{a,b}$ becomes small as well. Finally, for $t> t_2$, the first sum becomes negligible and in the limit $t\rightarrow\infty$, the system is in the coupled equilibrium $\omega_{\beta,\sigma,\lambda}$. In the degenerate situation, $\sigma=0$, the remainder term is small again after times $t> t_0$, and again after times $t> t_1$, the second sum in \fer{1} is negligible. However, since $\varepsilon_a(0,\lambda)=0$, the first sum is independent of time and does not decay. The initial state $\omega_0$ (applied to $\fm_0$) converges thus to the final state $\omega_\infty=\omega_{\beta,0,\lambda}+\sum_{a\geq 2}\chi_a$. The final state $\omega_\infty$ depends on the initial state $\omega_0$. It belongs to the manifold of $\alpha^t_{0,\lambda}$-invariant states on $\fm$, i.e., it is a convex combination $\sum_a\mu_a(\omega_0)\,\omega_{\s,a}\otimes\omega_{\r,a}$, with initial state dependent mixing parameters $\mu_a$.

Therefore, two time-scales emerge for the dynamics of systems in the overlapping resonances regime. On a time-scale $t_1\propto \lambda^{-2}$, which is very short with respect to $t_2
\propto \lambda^2/\sigma^2$, the initial state approaches a quasi-stationary manifold given by the first two terms on the r.h.s. of \fer{1}. For $\sigma=0$, this manifold is exactly stationary, but for $\sigma>0$ small, the manifold is only approximately stationary and it decays (into a the single equilibrium) for times exceeding $t_2\propto\lambda^2/\sigma^2$.

The appearence of different time-scales in open systems (albeit in somewhat different situations) has been observed before. The paper \cite{Dnew} examines the dynamics of a particle attracted by two widely separated potential wells and interacting with an infinite reservoir. The spacing of the wells, $1/\mu$, and the particle-reservoir interaction $\lambda$ are related by $\mu=\lambda^\beta$. It is shown that the dynamics of the particle in the weak coupling limit exists. The interaction between the wells has no effect for times of order $1/\lambda^2$ for $\beta>2$. However, for $0<\beta<2$ it has a direct effect on the particle dynamics and modifies the decay induced by the reservoir alone. The set of invariant states in the two regimes for $\beta$ are different.
In \cite{A}, various master equations for the dynamics of a nonlinear oscillator interacting with a reservoir are investigated. It is found that different generators yield more accurate descriptions of the reduced oscillator dynamics for different time-scales. In particular, different generators should be used for times shorter than, and longer than, the inverse of the system level-spacing. We mention that our analysis is valid for the {\em total} system-reservoir dynamics and for {\em all} times $t\geq 0$.

\medskip

{\bf Reduced dynamics.\ } Consider initial states of the form $\omega_0=\omega_{\s,0}\otimes\omega_{\r,\beta}$, where $\omega_{\s,0}$ is a state given by an arbitrary density matrix $\rho_0$ on ${\cx}^N$, $\omega_{\s,0}(A)={\rm Tr}_\s(\rho_0 A)$. The {\em reduced density matrix} $\rho_t$ of the system at time $t\geq0$ is defined by 
$$
{\rm Tr}_\s(\rho_t A)  = \omega_{\s,0}\otimes\omega_{\r,\beta}\big(\alpha^t_{\sigma,\lambda}(A)\big), \qquad\forall A\in{\cal B}({\cx}^N),
$$
where the trace is taken over the system space ${\cx}^N$. We denote the reduced evolution of the system by
$$
T_{\sigma,\lambda}(t)\rho_0 = \rho_t,
$$
and the manifold of initial system states which are invariant under the evolution, by
$$
{\cal M}_{\sigma,\lambda}=\{\rho_0\ :\ T_{\sigma,\lambda}(t)\rho_0= \rho_0 \mbox{\quad $\forall t\geq 0$}  \}.
$$
{}For $\sigma=0$ one can find the dynamics of the reduced density matrix exactly \cite{PSE,MSB,Privman} (see \fer{m31}). The manifold ${\cal M}_{0,\lambda}$ is the set of all system density matrices which are {\em diagonal in the eigenbasis of the interaction operator $G$}.  Moreover, we show in Appendix A that there is a constant $C$ such that, for all initial system states $\rho_0$ and all times $t\geq 0$,
\begin{equation}
\label{*}
{\rm dist}\big( {\cal M}_{0,\lambda}, T_{0,\lambda}(t)\rho_0 \big)  \leq C \e^{-\lambda^2\gamma_G\Gamma(t)}  {\rm dist}\big( {\cal M}_{0,\lambda}, \rho_0 \big).
\end{equation}
The distance ${\rm dist}( {\cal M}_{0,\lambda}, \rho)=\inf\{\|\tau-\rho\|_1\ :\ \tau\in{\cal M}_{0,\lambda}\}$ is measured in trace norm, $\|x\|_1={\rm Tr}\sqrt{xx^*}$ for linear operators $x$ on $\cx^N$.  Here, $\Gamma(t)\geq 0$ is the {\em decoherence function} (see \fer{m31.1}) and $\gamma_G=\min\{(g_a-g_b)^2 : a\neq b\}$, where $\{g_a\}_{a=1}^N$ is the spectrum  of $G$. Relation \fer{*} shows that  
the manifold ${\cal M}_{0,\lambda}$ is {\em orbitally stable}, meaning that a state initially close to ${\cal M}_{0,\lambda}$ remains so for all times.  If $\gamma_G>0$ and $\Gamma(t)\rightarrow\infty$ as $t\rightarrow\infty$, then the system undergoes full decoherence in the eigenbasis of $G$ (off-diagonal density matrix elements converge to zero as $t\rightarrow\infty$). In this case, \fer{*} shows that the manifold ${\cal M}_{0,\lambda}$ is dynamically attractive, or {\em asymptotically stable}. One shows that for suitable infra-red behaviour of the interaction form factor $g(k)$, the decoherence function satisfies  $\lim_{t\rightarrow\infty}\Gamma(t)/t=\Gamma_\infty$, with $\Gamma_\infty>0$.  The manifold ${\cal M}_{0,\lambda}$ is then approached exponentially quickly, at the rate $\lambda^2\gamma_G\Gamma_\infty$. We give further detail in Appendix \ref{appb}.

As the degeneracy is lifted, for small $\sigma>0$, the manifold of invariant initial system states becomes empty, ${\cal M}_{\sigma,\lambda}=\emptyset$. All initial states approach a single asymptotic state, which is the reduction to the small system of the joint system-reservoir equilibrium state (which is not a product state, see Appendix \ref{appb}). In the regime $\sigma<\!\!<|\lambda|<\!\!<1$, the approach of the asymptotic state, and hence the dissolution of the manifold ${\cal M}_{0,\lambda}$, takes place at a rate proportional to $\sigma^2/\lambda^2$, as we now show.  

The density matrix elements of the small system are given by
\begin{equation}
{}[\rho_t]_{a,b}\equiv\scalprod{\varphi_a}{\rho_t\,\varphi_b}, \qquad a,b=1,\ldots,N.
\label{dmatel}
\end{equation}
\begin{thm}[Reduced dynamics] 
\label{thmreddynfinal} 
Assume A1-A4. There is a $\lambda_0>0$ such that for fixed $\lambda$ satisfying $0<|\lambda|<\lambda_0$, the following holds. There is a $\sigma_0>0$ (depending on $\lambda$) s.t. if $0\leq\sigma<\sigma_0$, then we have, uniformly in $t\geq 0$:

-- For $a,b=1,\ldots,N$, $a\neq b$,
\begin{equation}
{}[\rho_t]_{a,b} = \e^{\i t\varepsilon_{b,a}(\sigma,\lambda)}[\rho_0]_{a,b} +O_\lambda(\sigma) +O(\lambda).
\label{l5}
\end{equation}

-- For $a=1,\ldots,N$,
\begin{equation}
{}[\rho_t]_{a,a} = \frac 1N+\sum_{b=2}^N D_{a,b}(t)[\rho_0]_{b,b} +O_\lambda(\sigma) +O(\lambda).
\label{l9}
\end{equation}
Let $\{\varphi^T_a\}_{a=1}^N$ be an orthonormal basis of eigenvectors of $T$, with $T\varphi_a^T=\xi_a\varphi_a^T$ and denote by $[\varphi^T_a]_c$, $c=1,\ldots,N$, the components of $\varphi^T_a$ (in the canonical basis). Then 
$$
D_{a,b}(t) =  \sum_{c=2}^N \e^{\i t\varepsilon_{c,c}(\sigma,\lambda)} \ \overline{[\varphi^T_c]_b} \ [\varphi^T_c]_{a}.
$$

\end{thm}

{\bf Discussion.\ } 1. The resonance energies governing the dynamics of {\em off-diagonals} are of the form (see \fer{l7}) 
$$
\varepsilon_{a,b}(\sigma,\lambda) = \lambda^2\delta_{a,b} +\sigma r_{a,b}+ \frac{\sigma^2}{\lambda^2}\,z_{a,b}+ O\left(\frac{\sigma^2}{\lambda}\right)+ O_\lambda(\sigma^3).
$$
We have the following interpretation:

$\bullet$  $\lambda^2\delta_{a,b}$ is a resonance energy for $\sigma=0$. The imaginary part of $\delta_{a,b}$ is proportional to $(g_a-g_b)^2$. All off-diagonal density matrix elements tend to zero (modulo an error term) as $t\rightarrow\infty$ if $g_a\neq g_b$ for $a\neq b$ and infra-red behaviour $p=-1/2$. The system exhibits then {\em decoherence in the eigenbasis of $G$}, regardless of whether the system energy is degenerate or not. The contribution to the decoherence rate of this term is proportional to $\lambda^2$.

$\bullet$ The term linear in $\sigma$ is real, with $r_{a,b}=[H_\s]_{a,a}-[H_\s]_{b,b}$. The decay rates of matrix elements do not depend on the first order in the energy splitting parameter $\sigma$.

$\bullet$ The second order term in $\sigma$ has generally non-vanishing real and imaginary parts. The complex $z_{a,b}$ are determined by the ratio of matrix elements $[H_\s]_{c ,d}$ and differences of $\delta_{c,d}$ (see \fer{pertevnon-zero}). The factor $1/\lambda^2$ is due to the presence of the reduced resolvent in second order perturbation theory in $\sigma$ (here, the `non-degenerate energies' are  $\lambda^2\delta_{a,b}$). The sign of ${\rm Im}\, z_{a,b}$ can be positive or negative, depending on the model.

2. The resonance energies driving the dynamics of the {\em diagonal} density matrix elements have the form
$$
\varepsilon_{c,c} (\sigma,\lambda)= 2\i\frac{\sigma^2}{\lambda^2} \xi_c + O\left(\frac{\sigma^2}{\lambda}\right)+ O_\lambda(\sigma^3).
$$
The $\xi_c$, $c=2,\ldots,N$, are strictly positive if, for instance, $[H_\s]_{a,b}{\rm Im}\delta_{a,b}\neq 0$ for all $a,b$ with $a\neq b$ (see after \fer{mmm5}). Then $D_{a,b}(t)$ decays exponentially quickly in time. Contrary to the off-diagonals, the diagonal entries of the density matrix evolve as a group: the value of a given diagonal entry depends on the initial condition of all of them. While the convergence rate of off-diagonals is proportional to $\lambda^2$, that of the diagonal is proportional to $\sigma^2/\lambda^2$. Hence the convergence of the diagonal, the part of the density matrix in the manifold ${\cal M}_{0,\lambda}$, is driven by the level splitting, while that of the off-diagonals is driven by the system-reservoir interaction.

\medskip

{\bf Transition between regimes for the spin-boson model. } We consider the small system to be a spin with Hamiltonian and interaction operator given by
\begin{equation*}
H_\s=S^z\equiv\frac{1}{2}\left(
\begin{array}{cc}
 1 & 0 \\
 0 & -1
\end{array}
\right) \quad \mbox{and}\quad G=S^x\equiv\frac{1}{2}\left(
\begin{array}{cc}
 0 & 1 \\
 1 & 0
\end{array}
\right),
\end{equation*}
respectively. The parameters $\sigma,\lambda$ are now considered to be small but independent of each other. We analyze the decoherence properties of the spin {\em in the energy basis}. Let $\phi^z_\pm$ be the normalized energy eigenvectors, satisfying $H_\s\phi^z_\pm=\pm\frac12\phi_\pm^z$, and denote the spin density matrix elements in this basis by $[\rho_t]^z_{+,-}:=\scalprod{\phi^z_+}{\rho_t\phi^z_-}$ (and similarly for other matrix elements). We show in Section \ref{sectspinboson} that  
\begin{eqnarray*}
{}[\rho_t]^z_{+,+}&\doteq&\textstyle\frac{1}{2}+\frac{1}{2}\e^{\i t
w_2}([\rho_0]^z_{+,+}-[\rho_0]^z_{-,-}),\\
{}[\rho_t]^z_{+,-}&\doteq& \textstyle\frac{r}{r^2+1}\left((1+r)\e^{\i tw_3}+(1/r-1)\e^{\i
tw_4}\right) [\rho_0]^z_{+,-},
\end{eqnarray*}
where $\doteq$ means that terms of order $O(\lambda^2)$ are disregarded (see \fer{u1}).  It is assumed here that $[\rho_0]^z_{+,-}\in\mathbb R$ (see \fer{1.106} for the general expression) and we have set
$$
r= \frac{-4\i \gamma -\sqrt{\pi^2\xi(0)^2 -16\gamma^2}}{\pi\xi(0)}\quad \mbox{with}\quad \gamma=\frac{\sigma}{\lambda^2}.
$$
Here, the square root is the principal branch with branch cut on the negative real axis and $\xi(0)>0$ is a constant proportional to the reservoir spectral density at zero (see \fer{xinot}). The system has four resonance energies, one is zero and the other three are
$$
w_2=\i \frac{\lambda^2}{2}\pi \xi(0), \quad w_{3,4}=\i\frac{\lambda^2}{4}\pi\xi(0)\pm\i \sqrt{\frac{\lambda^4}{16}\pi^2\xi(0)^2-\sigma^2}.
$$
These expressions interpolate the values of the previously known, isolated regime (lowest order in $\lambda$ for $\sigma$ fixed) and the overlapping resonances values derived here ($\sigma$ small, $\lambda$ fixed; see also the remark after Theorem \ref{thm4}).

The diagonal converges to $\frac12$ at the rate ${\rm Im}w_2\propto\lambda^2$, independently of $\sigma$. The decoherence rate (decay of the off-diagonal in the energy basis) is obtained as follows.
\begin{itemize}
\item[-] {\em Overlapping resonances regime}: $\gamma<\!\!<1$ and $r\approx -1$. Thus, $[\rho_t]^z_{+,-}\approx \e^{\i tw_4}[\rho_0]^z_{+,-}$, which has decay rate ${\rm Im}w_4\approx \frac{2}{\pi\xi(0)}\frac{\sigma^2}{\lambda^2}$.
\item[-] {\em Isolated resonances regime}: $1/\gamma<\!\!<1$ and  $r\approx-\i\infty$. Thus, $[\rho_t]^z_{+,-}\approx \e^{\i tw_3}[\rho_0]^z_{+,-}$, which has decay rate ${\rm Im}w_3\approx \frac{\pi\xi(0)}{4}\lambda^2$.
\end{itemize}
In the isolated resonances regime, the decoherence rate is given by the system-reservoir coupling constant $\lambda$ alone, while in the overlapping case, it depends also on the level splitting parameter $\sigma$. For a fixed $\lambda$, the decoherence rate increases quadratically in $\sigma$ (for small $\sigma$). The further its energy levels lie apart, the quicker the spin decoheres.

We define the critical value $\gamma_*$ for which the square root in $w_{3,4}$ vanishes,
$$
\gamma_* := \textstyle\frac14 \pi\xi(0).
$$
This critical value separates two regimes with different qualitative behaviour of the resonances $w_3$ and $w_4$. As $\gamma$ increases from zero to $\gamma_*$, the resonance $w_3$ moves down the imaginary axis, decreasing from the initial value $\frac12\i\pi\xi(0)\lambda^2$ to $\frac14\i\pi\xi(0)\lambda^2$, while $w_4$ moves up the imaginary axis, from the origin to $\frac14\i\pi\xi(0)\lambda^2$. The two resonances meet for $\gamma=\gamma_*$. As $\gamma>\gamma_*$ increases further, the resonances  $w_3$ and $w_4$ move horizontally away from the imaginary axis, their imaginary parts stay constant, equal to $\frac14\pi\xi(0)\lambda^2$. This motivates the sharp definition of the overlapping resonances regime, in the spin-boson model, to be given by $\gamma<\gamma_*$ and of the isolated resonances regime to be given by $\gamma>\gamma_*$. 

It is interesting to note that in nuclear physics, there is a (to our knowledge not rigorously defined) notion of overlapping resonances, used in the description of processes involving unstable nuclei by non-hermitian Hamiltonians \cite{SZ,CIZB}. It is observed that in the overlapping regime, the resonance widths (imaginary parts of resonance energies) segregate into two clusters, one located close to the origin (slow channels), the other at a much larger value (fast channels). The same occurs in our system: in the overlapping regime, we have one resonance at zero and another one, $w_4$, close to it. The other two, $w_2$ and $w_3$, are much larger, both close to $\frac12\i\pi\xi(0)\lambda^2$. As the system transitions into the isolated resonances regime, the two clusters mix.

\section{Resonances and dynamics}

\subsection{Resolvent representation}

The main result of this section is Theorem \ref{thm2}. For $\theta\in{\mathbb R}$ let $U_\theta$ be the unitary (translation) on $\cH_\r$ defined by sector-wise action $U_\theta \Omega_\r=\Omega_\r$ and $U_\theta\psi_n(u_1,\vartheta_1,\cdots,u_n,\vartheta_n)=\psi_n(u_1+\theta,\vartheta_1,\cdots,u_n+\theta,\vartheta_n)$. A vector $\psi\in\cH_\r$ is called $U_\theta$-analytic if the map $\theta\mapsto U_\theta\psi$ is $\cH_\r$-valued analytic in $\{\theta\in{\mathbb C} :\ |\theta|<\theta_0\}$ (the $\theta_0$ is that of assumption A1). All vectors of the form $\psi\otimes\Omega_\r$, for arbitrary $\psi\in\cH_\s$, are $U_\theta$-analytic. We introduce the {\em reference state}
\begin{equation}
\Omega=\Omega_\s\otimes\Omega_\r,
\label{m20}
\end{equation}
where $\Omega_\r$ is the vacuum in $\cH_\r$ and $\Omega_\s$ is the trace state 
\begin{equation}
\Omega_\s = \frac{1}{\sqrt{N}}\sum_{a=1}^N\varphi_a\otimes\varphi_a.
\label{m-1}
\end{equation}
$\Omega$ is cyclic and separating for $\fm$ and we denote the associated modular operator and modular conjugation by $\Delta$ and $J$, respectively \cite{BR}. We have $\Delta=\Delta_\s\otimes\Delta_\r$, where $\Delta_\r=\e^{-\beta L_\r}$ and $\Delta_\s=\bbbone$ (the trace state is KMS with inverse temperature $\beta=0$). The modular conjugation is $J=J_\s\otimes J_\r$. We have $J_\s\phi\otimes\chi = \bar\chi\otimes\bar\phi$ for $\phi,\chi\in{\mathbb C}^N$, and where the bar means complex conjugation of vector components in the basis $\{\varphi_a\}_{a=1}^N$. Furthermore, $J_\r\psi_n(u_1,\vartheta_1,\ldots,u_n,\vartheta_n) = \overline{\psi_n(-u_1,\vartheta_1,\ldots,-u_n,\vartheta_n)}$. A suitable generator of the dynamics is constructed as follows, see \cite{JP1} and also \cite{MMS}. On the dense set $\fm\,\Omega$ we define the group ${\cal U}(t)$ by
\begin{equation}
{\cal U}(t) A\Omega = \e^{\i tL(\sigma,\lambda)} A\e^{-\i tL(\sigma,\lambda)}\Omega,\qquad A\in\fm, \ t\in\rx,
\label{mm1}
\end{equation}
where $L(\sigma,\lambda)$ is the Liouvillian \fer{m11'}. We introduce the linear space
\begin{equation}
{\cal D}_0 = \dom(L_\r)\cap\dom(N^{1/2})\cap{\frak M}\, \Omega \subset{\cal H},
\label{m1}
\end{equation}
where $N=\d\Gamma(\bbbone)$ is the number operator. 
\begin{prop}
\label{propone}
{\bf (a)} ${\cal U}(t)$ is strongly differentiable on ${\cal D}_0$ and its
generator is given by
\begin{equation}
\i\frac{\d}{\d t}|_{t=0} \ {\cal U}(t) = K(\sigma,\lambda) := L_0(\sigma)+\lambda V -\lambda J\Delta^{1/2} V J\Delta^{1/2}.
\label{mm0}
\end{equation}
{\bf (b)} $\theta\mapsto U_\theta K(\sigma,\lambda) U_\theta^*$ has an analytic continuation from $\theta\in\rx$ to $\{\theta\in{\mathbb C}\ :\ |\theta|<\theta_0\}$,
in the strong sense on $\cD_0$. This continuation is given by
\begin{equation}
K_\theta(\sigma,\lambda) = L_{0,\theta}(\sigma)+\lambda I_\theta,
\label{b1}
\end{equation}
where
\begin{eqnarray}
L_{0,\theta}(\sigma) &=& L_0(\sigma)+\theta N\\
I_\theta&=& V_\theta -V'_\theta\\
V_\theta&=&\frac{1}{\sqrt 2}G\otimes
\bbbone\otimes\Big(a^*\big(g_\beta(\cdot+\theta)\big)+a\big(g_\beta(\cdot+\bar\theta)\big)\Big)\\
V'_\theta&=&\frac{1}{\sqrt 2}\bbbone\otimes
G\otimes\Big(a^*\big(e^{\frac{\beta}{2}(\cdot+\theta)}\overline{g}_\beta(-\cdot-\bar\theta)\big)+a\big(e^{-\frac{\beta}{2}(\cdot+\bar\theta)}
\overline{g}_\beta(-\cdot-\theta)\big)\Big)\qquad \label{b2}
\end{eqnarray}
(Here, we use the convention $\overline g(u)=\overline{g(u)}$.)
\end{prop}

\bigskip
\noindent
{\bf Proof.\ } We do not write the dependence of operators on $(\sigma,\lambda)$ in this proof, which follows \cite{JP1} (see also \cite{MMS}). 

{\bf (a)} Let $A\Omega\in\cD_0$. Then
\begin{equation}
\frac{\d}{\d t}|_{t=0} \ {\cal U}(t) A\Omega =-\i AL\Omega+\i LA\Omega =-\i A(L_0+\lambda V)\Omega+\i (L_0+\lambda V)A\Omega.
\label{m24}
\end{equation}
Since $L_0\Omega=0$ and $AV\Omega= J\Delta^{1/2}V^*A^*\Omega= J\Delta^{1/2}VJ\Delta^{1/2}A\Omega$, the right side of \fer{m24} equals $\i L_0A\Omega+ \i\lambda (V-J\Delta^{1/2} VJ\Delta^{1/2})A\Omega$. This shows part (a).

\smallskip
{\bf (b)} For real $\theta$, we have 
\begin{equation*}
\begin{split}
U_\theta K(\lambda)U_\theta^*=&L_0+\theta N+\frac{\lambda}{\sqrt
2}G\otimes
\bbbone\otimes\Big(a^*(g_\beta(\cdot+\theta))+a(g_\beta(\cdot+\theta))\Big)\\
-&\frac{\lambda}{\sqrt 2}\bbbone\otimes
G\otimes\Big(a^*(e^{\frac{\beta}{2}(\cdot+\theta)}\overline{g}_\beta(-\cdot-\theta))+a(e^{-\frac{\beta}{2}(\cdot+\theta)}\overline{g}_\beta(-\cdot-\theta))\Big).
\end{split}
\end{equation*}
By  assumption (A) we obtain the analytic extension \fer{b1}-\fer{b2}. Note that in the argument of the annihilation operators, the analytic extension has the complex conjugate $\bar\theta$, since the annihilation operators are anti-linear in their argument. \qed

\begin{thm}
\label{thm2} 
Assume A1 and A2. Let $\theta$ with $0<{\rm Im} \theta<\theta_0$ be fixed. There is a $\lambda_0>0$ such that for all $|\lambda|<\lambda_0$ and all $\sigma\in\mathbb R$, we have the following. Let $\phi\in {\cal H}$ and $A\in{\frak M}$ be such that $\phi$ and $A\Omega$ are $U_\theta$-analytic vectors, and such that $\phi_{\bar\theta}\in\dom(|L_\r|^{\frac14+\eta})$, for some $\eta>0$. Then we have for all $t\geq 0$
\begin{equation}
\scalprod{\phi}{\e^{\i tL(\sigma,\lambda)}A\e^{-\i tL(\sigma,\lambda)}\Omega} =
\frac{-1}{2\pi\i}\int_{\rx-\i} \e^{\i
tz}\scalprod{\phi_{\overline\theta}}{(K_\theta(\sigma,\lambda)-z)^{-1}
(A\Omega)_\theta}\d z. 
\label{mm2}
\end{equation}
\end{thm}
We give a proof of this result in Appendix \ref{appa}. 

{\bf Remarks.\ } 1. Vectors representing product states of an arbitrary small system state and the equilibrium reservoir states are of the form $\phi=B\Omega$, where $B\in\fm_\s$ (and, recall, $\Omega$ is given in \fer{m20}). The proof of \fer{mm2} for such $\phi$ and $A\in\fm_\s$ is easier than that of the full result. This is the situation of \cite{MSB}.

2. In \cite{MMS} a spectral dilation deformation is performed simultaneously with the translation (see also \cite{BFS,MMS1}). In this doubly-deformed situation, the analogue of Theorem \ref{thm2} is proven in Section 8 of \cite{MMS}. The dilation deforms the spectrum of $K$ in a `sectorial way' (a $V$-shape), leading to useful decay estimates of the (deformed) resolvent $(K-z)^{-n}$, as $|{\rm Re}z|\rightarrow\infty$. However, in the present work, we only use spectral translation and such decay estimates do not hold (as the distance between the spectrum of $K_\theta$ and the real axis does not grow now when $|{\rm Re} z|\rightarrow\infty$). We therefore need a new proof of this result. The advantage of only performing the translation deformation is that less restrictive conditions on the form factor are needed only.

\subsection{Resonances of $K(\sigma=0,\lambda)$}

The operator $K_\theta(0,\lambda)$ is defined in Proposition \ref{propone}, with $L_0=L_\r$. Recall that $\varphi_a$, $a=1,\ldots, N$, is the orthonormal eigenbasis of $G$, \fer{m22}. The operator $K_\theta(0,\lambda)$ is reduced by the decomposition 
$$
\cH = \bigoplus_{a,b=1}^N {\, \rm Ran\,}\Big(|\varphi_a\rangle\langle \varphi_a|\otimes|\varphi_b\rangle\langle \varphi_b|\Big)\otimes\cH_\r.
$$
Namely,
\begin{equation}
K_\theta(0,\lambda) = \bigoplus_{a,b=1}^N K_{a,b},
\label{kab}
\end{equation}
where $K_{a,b}$ acts on $\cH_\r$ as
\begin{equation}
K_{a,b}=L_\r+\theta N+\lambda(g_a\Phi_\theta-g_b\widetilde\Phi_\theta),
\label{01.33}
\end{equation}
with
\begin{equation}
\begin{split}
\Phi_\theta=&\frac{1}{\sqrt 2}\big( a^*(g_\beta(\cdot+\theta))+a(g_\beta(\cdot+\bar\theta))\big)\\
\widetilde \Phi_\theta=&\frac{1}{\sqrt
2}\big(a^*(\e^{\frac{\beta}{2}(\cdot+\theta)}\overline{g}_\beta(-\cdot-\theta))+a(\e^{-\frac{\beta}{2}(\cdot+\bar\theta)}\overline{g}_\beta(-\cdot-\bar\theta))\big).
\end{split}
\end{equation}
To alleviate the notation, we do not display $\theta$ and $\lambda$ in $K_{a,b}$.
\begin{thm}[Spectrum of $K_{a,b}$]
\label{thm3}
Assume A1 and A2. Let $\theta$ with $0<{\rm Im \theta}<\theta_0$ be fixed. 
There is a $\lambda_0>0$ such that if $0\leq |\lambda|<\lambda_0$, then for all $a,b=1,\ldots,N$, the operator $K_{a,b}$ has a simple eigenvalue $\lambda^2\delta_{a,b}$, where $\delta_{a,b}$ is given in \fer{delta}. All other spectrum of $K_{a,b}$ lies in $\{z\in\cx\ :\ {\rm Im}z>\frac34{\rm Im}\theta\}$.
\end{thm}

{\bf Remarks.\ } 1. It follows from Theorem \ref{thm3} and the decomposition \fer{kab} that the spectrum of $K_\theta(0,\lambda)$ in the strip $\{z\in\cx\ :\ {\rm Im}z<\frac34{\rm Im}\theta\}$ consists precisely of the eigenvalues $\{\lambda^2\delta_{a,b}\}_{a,b=1}^N$ (there are no higher order terms in $\lambda$). A simple expression for the eigenvectors associated to the non-zero eigenvalues is not available, only a perturbation series is. However, it is readily seen that the eigenvalue zero has the eigenvectors $\varphi_a\otimes\varphi_a\otimes\Omega_\r$, $a=1,\ldots,N$. Indeed, if $a=b$, then it follows directly from \fer{01.33} that 
\begin{equation}
K_{a,a}\Omega_\r =\lambda g_a U_\theta (\Phi-J\Delta^{1/2} \Phi J\Delta^{1/2})\Omega_\r =0,
\label{remark}
\end{equation}
since $J\Delta^{1/2}\Phi J\Delta^{1/2}\Omega_\r=\Phi\Omega_\r$.

2. If the form factor $g$ satisfies $\|g_\beta/u\|^2_2<\infty$, then the operator $K_{a,b}$, \fer{01.33}, is unitarily equivalent to the operator $L_\r+{\rm const.}$ The condition on the form factor implies the infra-red behaviour $g(k)\sim|k|^p$ for small $k$, with $p>-1/2$. Then $K_{a,b}$ has a simple real eigenvalue, as also predicted by \fer{delta}, saying that ${\rm Im}\delta_{a,b}=0$. In the infra-red singular case, $p=-1/2$, the unitary transformation ceases to exist and the eigenvalue becomes complex.

\medskip

\noindent
{\bf Proof of Theorem \ref{thm3}.\ } The spectrum of $K_{a,b}$ for $\lambda=0$ consists of a single simple eigenvalue at zero, with eigenvector $\Omega_\r$, and of horizontal lines of continuous spectrum $\{ x+{\rm Im} \theta \,n\ :\ x\in{\mathbb R}, n=1,2,\ldots\}$. The operators $\Phi_\theta$ and $\widetilde\Phi_\theta$ are infinitesimally small w.r.t. $N$ (relatively bounded with arbitrarily small relative bound). Analytic perturbation theory implies that there exists a $\lambda_0>0$ such that if $0\leq |\lambda|<\lambda_0$, then the only spectrum of $K_{a,b}$ in $\{ z\in{\mathbb C}\ :\ {\rm Im}z<{\rm Im}\theta/2\}$ is a single, simple eigenvalue. We show that this eigenvalue is $\lambda^2\delta_{a,b}$, with $\delta_{a,b}$ given in \fer{delta}.

The dynamics of the {\em reduced density matrix} of the small system has been calculated explicitly in Proposition 7.4 of \cite{MSB}. Let $\psi_0=B\Omega_\s\otimes \Omega_\r$ be an initial state, where $B\in\fm_\s'$ (the commutant) is arbitrary (see also \fer{m20}). The reduced system density matrix at time $t$, in the basis $\{\varphi_a\}$, is given by $[\rho_t]_{a,b}= \langle \psi_0, \e^{\i tL(0,\lambda)}(|\varphi_b\rangle \langle\varphi_a|\otimes \bbbone_\s) \e^{-\i tL(0,\lambda)}\psi_0\rangle$. 
It is shown in the above reference that 
\begin{equation}
[\rho_t]_{a,b}=[\rho_0]_{a,b}\ \e^{\i\lambda^2\alpha_{a,b}(t)},
\label{m31}
\end{equation}
with
$\alpha_{a,b}(t)=(g_a^2-g_b^2)S(t)+\i(g_a-g_b)^2\Gamma(t)$, where
\begin{equation}
\Gamma(t)=\textstyle\int_{{\mathbb R}^3}|g(k)|^2\coth
(\frac{\beta|k|}{2})\frac{\sin ^2(\frac{|k|
t}{2})}{|k|^2}\d^3k,\quad
S(t)=\frac{1}{2}\int_{{\mathbb R}^3}|g(k)|^2\frac{|k| t-\sin
|k| t}{|k|^2}\d^3k.
\label{m31.1}
\end{equation}
For large times, $\alpha_{a,b}(t)$ becomes linear,
\begin{equation}
 \lim_{t\rightarrow\infty}\frac{\alpha_{a,b}(t)}{t}=\delta_{a,b},
\label{m32}
\end{equation}
with $\delta_{a,b}$ given in \fer{delta}. We express the reduced density matrix alternatively, using Theorem \ref{thm2}, as
\begin{equation}
[\rho_t]_{a,b}=\frac{-1}{2\pi \i}\int_{ \mathbb{R}-\i}\e^{\i tz}\scalprod{
B^*B\Omega_\s\otimes\Omega_\r}{(K_\theta-z)^{-1}\big(|\varphi_b\rangle
\langle\varphi_a|\otimes \bbbone_\s\big) \Omega_\s\otimes\Omega_\r} \d z.
\label{m33}
\end{equation}
We use that $\e^{\i tL(0,\lambda)}(|\varphi_b\rangle \langle\varphi_a|\otimes \bbbone_\s) \e^{-\i tL(0,\lambda)}B = B\e^{\i tL(0,\lambda)}(|\varphi_b\rangle \langle\varphi_a|\otimes \bbbone_\s) \e^{-\i tL(0,\lambda)}$, which holds since $B\otimes\bbbone_\r$ belongs to the commutant $\fm'$. It follows from the definition \fer{m-1} that $(|\varphi_b\rangle \langle\varphi_a|\otimes \bbbone_\s)\Omega_S=\frac{1}{\sqrt{N}}\varphi_b\otimes \varphi_a$. Therefore, we obtain from \fer{m33} that
\begin{eqnarray}
[\rho_t]_{a,b} &=&\frac{1}{\sqrt N}\scalprod{B^*B\Omega_\s}{\varphi_b\otimes \varphi_a}\ \frac{-1}{2\pi \i}\int_{\mathbb R-\i}\e^{\i tz}\scalprod{\Omega_\r}{(K_{b,a}-z)^{-1}\Omega_\r} \d z\nonumber\\
&=&[\rho_0]_{a,b}\ \frac{-1}{2\pi \i}\int_{\R-\i}\e^{\i tz}\scalprod{\Omega_\r}{(K_{b,a}-z)^{-1}\Omega_\r}\d z.
\label{01.16}
\end{eqnarray}
Comparing \fer{01.16} and \fer{m31} yields the identity
\begin{equation}
\e^{\i\lambda^2\alpha_{a,b}(t)}=\frac{-1}{2\pi \i}\int_{\mathbb
R-\i}\e^{\i tz}\scalprod{\Omega_\r}{(K_{b,a}-z)^{-1}\Omega_\r}\d z.
\label{01.39}
\end{equation}
Denote the unique eigenvalue of $K_{a,b}$ in $\{z\in\cx\ :\ {\rm Im} z< {\rm Im}\theta/2\}$ by $\zeta_{a,b}(\lambda)$ and let ${\mathcal C}_{a,b}$ be a small circle around $\zeta_{a,b}(\lambda)$ not including any other point of the spectrum of $K_{a,b}$. By deforming the contour of integration, we have
\begin{equation}
\frac{-1}{2\pi \i}\int_{\mathbb R-\i}\e^{\i tz}\scalprod{\Omega_\r}{(K_{a,b}-z)^{-1}\Omega_\r}\d z = \frac{-1}{2\pi
\i}\oint_{{\mathcal C}_{a,b}}\e^{\i tz}\scalprod{\Omega_\r}{(K_{a,b}-z)^{-1}\Omega_\r}\d z +R_\lambda(t),
\label{01.40}
\end{equation}
with a remainder term small in $\lambda$ and decaying to zero exponentially quickly as $t\rightarrow\infty$. This follows from the following result, proven in \cite{MSB}, Proposition 4.2:
\begin{prop}[\cite{MSB}]
\label{prop7}
Let $\psi_0\in\cH_\s$. Then
$$
\left|\int_{\rx+\i\frac34{\rm Im}\theta}\e^{\i t z}\scalprod{\psi_0\otimes\Omega_\r}{(K_\theta(\sigma,\lambda)-z)^{-1}\psi_0\otimes\Omega_\r}\d z\right|\leq C \lambda^2 \e^{-\frac34t\,{\rm Im\theta}},
$$ 
uniformly in $\sigma$ varying in compact sets. The same bound holds if $K_\theta(\sigma,\lambda)$ is replaced by $K_{a,b}$.
\end{prop}
This result implies that $|R_\lambda(t)|\leq C\lambda^2\e^{-\frac{3{\rm Im \theta}}{4}t}$ for some constant $C$.
Since $\zeta_{a,b}(\lambda)$ is a simple pole of the resolvent $(K_{a,b}-z)^{-1}$ we can replace $\e^{\i tz}$ by $\e^{\i t \zeta_{a,b}(\lambda)}$ in \fer{01.40} and we obtain
\begin{equation}
\frac{-1}{2\pi \i}\int_{\R-\i}\e^{\i tz}\scalprod{\Omega_\r}{(K_{a,b}-z)^{-1}\Omega_\r}\d z = \e^{\i t\zeta_{a,b}(\lambda)} c_{a,b}(\lambda)+R_\lambda(t),
\label{m35}
\end{equation}
where $c_{a,b}(\lambda)=\frac{-1}{2\pi
\i}\oint_{{\mathcal{C}}_{a,b}}\scalprod{\Omega_\r}{(K_{a,b}-z)^{-1}\Omega_\r}\d z$. Combining \fer{01.39} and \fer{m35} gives 
$$
 \e^{\i\lambda^2\alpha_{a,b}(t)-\i t\zeta_{a,b}(\lambda)} = c_{a,b}(\lambda)+\e^{-\i t\zeta_{a,b}(\lambda)} R_\lambda(t).
$$
As ${\rm Im}\zeta_{a,b}(\lambda)<\frac12{\rm Im}\theta$, we have $\lim_{t\rightarrow\infty}\e^{-\i t\zeta_{a,b}(\lambda)} R_\lambda(t)=0$. Thus the exponent on the left hand side converges to a finite number, as $t\rightarrow\infty$, and so this exponent, divided by $t$, tends to zero as $t\rightarrow\infty$. (Note that $c_{a,b}(\lambda)$ is not zero for small $\lambda$, by perturbation theory.) Then, due to  \fer{m32}, we have $\zeta_{a,b}(\lambda)=\lambda^2\delta_{a,b}$. The proof of Theorem \ref{thm3} is complete. \hfill $\square$

\subsection{Resonances of $K(\sigma,\lambda)$}

We now examine the operator $K_\theta(\sigma,\lambda)$, defined in Proposition \ref{propone}, \fer{b1}-\fer{b2}, with $L_0$ given in \fer{m12}. We consider  $K_\theta(\sigma,\lambda)$ as an unperturbed part, $K_\theta(0,\lambda)$, plus a perturbation $\sigma L_\s$ (see \fer{m11}). Since the eigenvalues of $K_\theta(0,\lambda)$ are isolated (Theorem \ref{thm3}), we can apply analytic perturbation theory to follow them as the perturbation is switched on ($\sigma\neq 0$).

\begin{thm}[Spectrum of $K_\theta(\sigma,\lambda)$]
\label{thm4}
Assume A1-A3. Let $\lambda$ be fixed, satisfying $0<|\lambda|<\lambda_0$, where $\lambda_0$ is given in Theorem \ref{thm3}. There is a $\sigma_0>0$ (depending on $\lambda$) s.t. if $0\leq \sigma<\sigma_0$, then the spectrum of $K_\theta(\sigma,\lambda)$ in the region $\{z\in\cx\ :\ {\rm Im}z<\frac12{\rm Im}\theta\}$ consists of simple eigenvalues $\varepsilon_{a,b}(\sigma,\lambda)$. Those eigenvalues are analytic functions of $\sigma$, given by \fer{l7}. Zero is an eigenvalue of $T$, \fer{mmm6}. It is simple if $[H_\s]_{a,b}\neq 0$ for all $a\neq b$.
\end{thm}

{\bf Remark. } The theorem assumes the non-degeneracy condition A3. An analysis in presence of degenerate non-zero resonances $\lambda^2\delta_{a,b}$ can be carried out along the same lines. We have done this for the spin-boson model. We have checked that the values for the resonances thus obtained coincide with those obtained in Section \ref{sectspinboson} (to order two in $\sigma$).

\medskip
\noindent
{\bf Proof of Theorem \ref{thm4}.\ } {\bf (A) Non-zero eigenvalues.} The non-zero eigenvalues of $K_\theta(0,\lambda)$ are simple, given by $\varepsilon_{a,b}(0,\lambda)=\lambda^2\delta_{a,b}$, for $a\neq b$. We denote by $\varphi_{a,b}\otimes X_{a,b}$ the eigenvector associated to $\varepsilon_{a,b}(0,\lambda)$, where $\varphi_{a,b}=\varphi_a\otimes\varphi_b$ and $X_{a,b}$ is a normalized vector in $\cH_\r$, depending on $\lambda$ and $\theta$. The adjoint operator satisfies $K_\theta(0,\lambda)^*\varphi_{a,b}\otimes X_{a,b}^*=\lambda^2\overline{\delta}_{a,b}\varphi_{a,b}\otimes X_{a,b}^*$ for a vector $X^*_{a,b}$ satisfying $\scalprod{X_{a,b}}{X^*_{a,b}}=1$. We denote the Riesz projection of $K_\theta(0,\lambda)$ associated to $\varepsilon_{a,b}(0,\lambda)$ by 
\begin{equation}
P_{a,b} = |\varphi_{a,b}\otimes X_{a,b}\rangle\langle \varphi_{a,b}\otimes X^*_{a,b}|.
\label{pab}
\end{equation}
By analytic perturbation theory, $K_\theta(\sigma,\lambda)$ has a simple eigenvalue in the vicinity of $\lambda^2\delta_{a,b}$, for small $\sigma$. It is given by 
\begin{equation}
\varepsilon_{a,b}(\sigma,\lambda) =\lambda^2\delta_{a,b}+\sigma \varepsilon_{a,b}^{(1)}+\sigma^2 \varepsilon_{a,b}^{(2)} +O_\lambda(\sigma^3),
\end{equation}
where (see \cite[Sect. II.2.2]{Kato} and also \cite[Thm. XII.12]{RS4}) 
\begin{equation}
\varepsilon_{a,b}^{(1)} = {\rm Tr} (L_\s P_{a,b}) = [H_\s]_{a,a}-[H_\s]_{b,b}.
\label{m50}
\end{equation}
Here, we have set $[H_\s]_{a,b}=\scalprod{\varphi_a}{H_\s\varphi_b}$. The second order correction is 
\begin{equation}
\varepsilon_{a,b}^{(2)}= -{\rm Tr} \big( L_\s (K_\theta(0,\lambda)-\lambda^2\delta_{a,b})^{-1}\bar P_{a,b} L_\s P_{a,b}\big).
\label{m51}
\end{equation}
We write $\bar P$ for $\bbbone-P$ for general projections $P$. We set $P^\s_{a,b}=|\varphi_{a,b}\rangle\langle\varphi_{a,b}|$ and $P^\r_{a,b}=|X_{a,b}\rangle\langle X_{a,b}^*|$. Then $P_{a,b}=P^\s_{a,b}\otimes P^\r_{a,b}$ and $\bar P_{a,b} = \bar P^\s_{a,b}\otimes\bbbone_\r + P^\s_{a,b}\otimes\bar P^\r_{a,b}$. It follows that 
$\bar P_{a,b} L_\s\ (\varphi_{a,b}\otimes X_{a,b}) = (\bar P^\s_{a,b} L_\s\varphi_{a,b})\otimes X_{a,b}$. 
Using this and $\bar P^\s_{a,b}=\sum_{(c,d)\neq(a,b)}P^\s_{c,d}$ in expression \fer{m51} yields
\begin{equation*}
\varepsilon_{a,b}^{(2)} =-\sum_{(c,d)\neq (a,b)}\scalprod{\varphi_{a,b}\otimes X_{a,b}^*}{L_\s\, (K_\theta(0,\lambda)-\lambda^2\delta_{a,b})^{-1} \varphi_{c,d}\otimes X_{a,b}} \scalprod{\varphi_{c,d}}{L_\s\varphi_{a,b}}.
\end{equation*}
`Replacing' $\varphi_{c,d}\otimes X_{a,b}$ by the eigenvector $\varphi_{c,d}\otimes X_{c,d}$, we obtain
\begin{equation}
\varepsilon_{a,b}^{(2)} =-\sum_{(c,d)\neq (a,b)}\frac{1}{\lambda^2(\delta_{c,d}-\delta_{a,b})} |\scalprod{\varphi_{a,b}}{L_\s \varphi_{c,d}}|^2 \scalprod{X^*_{a,b}}{X_{c,d}} +\xi,
\label{mmm1}
\end{equation}
where 
\begin{equation}
\xi =\sum_{(c,d)\neq (a,b)} \scalprod{\varphi_{a,b}\otimes X^*_{a,b}}{L_\s (K_\theta(0,\lambda)-\lambda^2\delta_{a,b})^{-1}\varphi_{c,d}\otimes(X_{c,d}-X_{a,b})}\scalprod{\varphi_{c,d}}{L_\s\varphi_{a,b}}.
\label{m51''}
\end{equation}
By perturbation theory, we have $X_{a,b}=\Omega_\r+O(\lambda)$. Therefore, $X_{c,d}-X_{a,b}=O(\lambda)$ and $\scalprod{X^*_{a,b}}{X_{c,d}}=1+O(\lambda)$. Together with the bound \fer{resbnd} of Corollary \ref{corollary1} below, we obtain
\begin{equation}
|\xi|\leq \frac{C}{|\lambda|}.
\label{m51.1}
\end{equation} 
Finally, 
\begin{equation}
\scalprod{\varphi_{a,b}}{L_\s\varphi_{c,d}} = \chi_{b=d}\, [H_\s]_{a,c}-\chi_{a=c}\, [H_\s]_{d,b}.
\label{bnd3}
\end{equation}
Relation \fer{l7} for $a\neq b$ follows from \fer{mmm1}, \fer{m51.1} and \fer{bnd3} and a little algebra. 
\begin{prop}[Bound on the resolvent]
\label{resbndprop}
There are constants $C$ and $\lambda_0$ (depending on ${\rm Im}\theta$ only) such that if $0<|\lambda|<\lambda_0$, then we have the following. Fix any $\alpha>0$ and take complex $z$ satisfying $|z|<C\alpha$, ${\rm Im}z<\frac14{\rm Im}\theta$, and ${\rm dist}({\cal E},z)\geq\alpha\lambda^2$, where ${\cal E}=\{\lambda^2\delta_{a,b}\, : \, a,b=1,\ldots,N\}$ is the set of eigenvalues of $K_\theta(0,\lambda)$. Then we have 
\begin{equation}
\|(K_\theta(0,\lambda)-z)^{-1}\|\leq C_1 \left(\frac{1}{{\rm Im}\theta} +\frac{1}{{\rm dist}({\cal E},z)}\right),
\label{resbnd1}
\end{equation}
where $C_1$ is a constant depending only on ${\rm Im}\theta$.
\end{prop}
Knowing the bound on the resolvent we can obtain a bound on the {\em reduced resolvent}.
\begin{cor}
\label{corollary1}
{}For any $a,b=1,\ldots,N$ we have 
\begin{equation}
\|(K_\theta(0,\lambda)-\lambda^2\delta_{a,b})^{-1}\bar P_{a,b}\|\leq C_2\left( \frac{1}{{\rm Im}\theta} + \frac{1}{\lambda^2}\right),
\label{resbnd}
\end{equation}
for some constant $C_2$ depending on ${\rm Im}\theta$.
\end{cor}

\noindent
{\bf Proof of Corollary \ref{corollary1}.\ } The reduced resolvent has the representation
$$
(K_\theta(0,\lambda)-\lambda^2\delta_{a,b})^{-1}\bar P_{a,b} = \frac{-1}{2\pi\i}\oint_{\Gamma_{a,b}(\lambda)} (z-\lambda^2\delta_{a,b})^{-1} (K_\theta(0,\lambda)-z)^{-1}\bar P_{a,b} \d z,
$$
where $\Gamma_{a,b}(\lambda)=\{z=\lambda^2\delta_{a,b}+\lambda^2 r\e^{\i\phi}\, :\, \phi\in[0,2\pi]\}$, with an appropriate radius $r$ (independent of $\lambda$) such that $\Gamma_{a,b}(\lambda)$ encircles only the eigenvalue $\lambda^2\delta_{a,b}$ and such that $\Gamma_{a,b}(\lambda)$ lies within the region of $z$ for which the bound \fer{resbnd1} holds, according to Proposition \ref{resbndprop}. Then ${\rm dist}({\cal E},z)$ is a constant times $\lambda^2$. It follows that 
$$
\|(K_\theta(0,\lambda)-\lambda^2\delta_{a,b})^{-1}\bar P_{a,b}\|\leq C \left(\frac{1}{{\rm Im}\theta} +\frac{1}{\lambda^2}\right)(1+\|P_{a,b}\|),
$$
for some constant $C$. The bound \fer{resbnd} follows from $\|P_{a,b}\|=1+O(\lambda)$. \hfill $\square$

\noindent
{\bf Proof of Proposition \ref{resbndprop}.\ }  Let $P_\r=|\Omega_\r\rangle\langle\Omega_\r|$, $\bar P_\r=\bbbone-P_\r$, and $R(z)=(K_\theta(0,\lambda)-z)^{-1}$.

{\em Step 1.} For any $\psi\in\cH$ we have 
\begin{eqnarray}
\lefteqn{
\left|\scalprod{\psi}{\bar P_\r (K_\theta(0,\lambda)-z)\bar P_\r\psi}\right|
\geq{\rm Im}\scalprod{\psi}{\bar P_\r (K_\theta(0,\lambda)-z)\bar P_\r\psi}}\nonumber\\
&=& \scalprod{\psi}{\bar P_\r\big(N^{1/2}\{ {\rm Im}\theta+ \lambda{\rm Im}N^{-1/2}I_\theta N^{-1/2}\}N^{1/2} -{\rm Im}z \big)\bar P_\r\psi}\nonumber\\
&\geq&({\rm Im}\theta -C |\lambda| - {\rm Im}z)\|\bar P_\r\psi\|^2\nonumber\\
&\geq&\textstyle\frac12{\rm Im}\theta\,\|\bar P_\r\psi\|^2.\nonumber
\end{eqnarray}
By the Cauchy-Schwartz inequality, it follows that $\|\bar P_\r(K_\theta(0,\lambda)-z)\bar P_\r\psi\|\geq \frac 12{\rm Im}\theta\,\|\bar P_\r\psi\|$ and therefore
\begin{equation}
\|\bar P_\r R(z) \bar P_\r\|\leq \frac{2}{{\rm Im}\theta}.
\label{bnd1}
\end{equation}

{\em Step 2.} Consider the Feshbach map 
\begin{eqnarray}
{\cal F}_z&=& P_\r(-z-\lambda^2 I_\theta \bar P_\r R(z)\bar P_\r I_\theta)P_\r\nonumber\\
&=&  P_\r(-z-\lambda^2 I_\theta \bar P_\r R(0)\bar P_\r I_\theta)P_\r +O(\lambda^2|z|).
\label{diffg}
\end{eqnarray}
Let
\begin{equation}
{\cal G}_z = -\lambda^2 P_\r I_\theta \bar P_\r R(z)\bar P_\r I_\theta P_\r.
\label{np1}
\end{equation}
By the isospectrality property of the Feshbach map (see e.g. \cite[Theorem IV.1]{BFS2}) we know that 
$$
{\cal G}_{\lambda^2\delta_{a,b}}\,\varphi_{a,b}\otimes\Omega_\r = \lambda^2\delta_{a,b} \,\varphi_{a,b}\otimes\Omega_\r,
$$
for all $a,b=1,\ldots,N$. We also have ${\cal G}_z-{\cal G}_\zeta = O(\lambda^2|z-\zeta|)$, as long as ${\rm Im}z, {\rm Im}\zeta <\frac14{\rm Im}\theta$. It follows that ${\cal G}_0\,\varphi_{a,b}\otimes\Omega_\r = \lambda^2\delta_{a,b}\,\varphi_{a,b}\otimes\Omega_\r +O(\lambda^4)$, for all $a,b=1,\ldots,N$. Therefore, ${\cal G}_0=\sum_{a,b=1}^N\lambda^2\delta_{a,b} |\varphi_{a,b}\rangle\langle\varphi_{a,b}|\otimes P_\r +O(\lambda^4)$,
and so
\begin{equation}
{\cal G}_z = \sum_{a,b=1}^N\lambda^2\delta_{a,b} |\varphi_{a,b}\rangle\langle\varphi_{a,b}|\otimes P_\r +O(\lambda^4+\lambda^2|z|).
\label{np2}
\end{equation}
Using \fer{np2} and \fer{np1} in \fer{diffg} shows that
\begin{equation}
{\cal F}_z = \sum_{a,b=1}^N (\lambda^2\delta_{a,b} -z) |\varphi_{a,b}\rangle\langle\varphi_{a,b}|\otimes P_\r +O(\lambda^4+\lambda^2|z|).
\label{np3}
\end{equation}
The sum on the right side is an invertible operator, the norm of the inverse being 
$$
\max_{a,b=1,\ldots,N}|\lambda^2\delta_{a,b}-z|^{-1}=[{\rm dist}({\cal E},z)]^{-1}.
$$
Therefore, there is a constant $C$ s.t. if 
\begin{equation}
\lambda^4+\lambda^2|z|< C\,{\rm dist}({\cal E},z),
\label{np5}
\end{equation}
then ${\cal F}_z$ is invertible and 
\begin{equation}
\|{\cal F}_z^{-1}\|\leq \frac{2}{{\rm dist}({\cal E},z)}.
\label{np4}
\end{equation}
Let $\alpha>0$ be fixed, and take $z$ s.t. ${\rm dist}({\cal E},z)\geq\alpha\lambda^2$. Then \fer{np5} is satisfied provided $\lambda$ is small enough and $|z|< C\alpha$.

\medskip

{\em Step 3.} The resolvent $R(z)$ is related to $\bar P_\r R(z)\bar P_\r$ and ${\cal F}_z^{-1}$ by (see e.g. \cite[Eqn. (IV.14)]{BFS2})
$$
R(z) = \big( P_\r -\bar P_\r R(z)\bar P_\r K_\theta(0,\lambda)P_\r\big){\cal F}_z^{-1} \big( P_\r - P_\r K_\theta(0,\lambda)\bar P_\r R(z)\bar P_\r\big) +\bar P_\r R(z)\bar P_\r.
$$
We combine this equation with the bounds $\|\bar P_\r K_\theta(0,\lambda)P_\r\|$, $\|P_\r K_\theta(0,\lambda)\bar P_\r\|\leq C|\lambda|$ and \fer{bnd1}, \fer{np4} to arrive at the estimate \fer{resbnd1}. This completes the proof of Proposition \ref{resbndprop}.\hfill $\square$

{\bf (B) Zero eigenvalue.} Let $P(\sigma)$ be the group projection associated to the eigenvalues of $K_\theta(\sigma,\lambda)$ bifurcating out of the origin as $\sigma\neq 0$. Here, we consider $\lambda$ fixed and $\sigma$ small. The null space of $K_\theta(0,\lambda)$ is known exactly, see \fer{remark}. Let $X^*_{a,a}\in\cH_\r$ be the vector satisfying $K^*_{a,a}X^*_{a,a}=0$ and $\scalprod{\Omega_\r}{X^*_{a,a}}=1$. We have $X^*_{a,a}=\Omega_\r+O(\lambda)$. Then $P(0)=\sum_{a=1}^N|\varphi_{a,a}\rangle\langle\varphi_{a,a}|\otimes|\Omega_\r\rangle\langle X^*_{a,a}|$. Note that $P(0) L_\s P(0)=0$. Analytic perturbation theory gives
\begin{eqnarray}
K_\theta(\sigma,\lambda) P(\sigma)&=& \sigma^2 T_2 +O_\lambda(\sigma^3)
\nonumber\label{mmm1'}\\
T_2&=& -P(0) L_\s  K_\theta(0,\lambda)^{-1} L_\s P(0).
\label{mmm2}
\end{eqnarray}
We have $ L_\s P(0) = \sum_{a=1}^N\sum_{c,d=1,\ldots,N; c\neq d} |\varphi_{c,d}\rangle\langle\varphi_{a,a}| \otimes P_\r\scalprod{\varphi_{c,d}}{ L_\s \varphi_{a,a}} +O(\lambda)$. Next,
\begin{eqnarray}
K_\theta(0,\lambda)^{-1} \varphi_{c,d}\otimes \Omega_\r &=& K_\theta(0,\lambda)^{-1} \varphi_{c,d}\otimes (X_{c,d}+\Omega_\r-X_{c,d})\nonumber\\
&=&  \frac{1}{\lambda^2\delta_{c,d}} \varphi_{c,d}\otimes \Omega_\r +O(\lambda^{-1}),
\label{mnm1}
\end{eqnarray}
where we use Corollary \ref{corollary1} in the last step. Starting from \fer{mmm2} and using \fer{mnm1}, we arrive at 
\begin{equation}
T_2 = \frac{2\i}{\lambda^2} T+O(\lambda^{-1}),
\label{mmm5}
\end{equation}
where the operator $T$ has matrix elements $[T]_{a,b}=\scalprod{\varphi_{a,a}\otimes\Omega_\r}{T\,\varphi_{b,b}\otimes\Omega_\r}$ given by \fer{mmm6}. In this derivation, we also use that $\delta_{b,a}=-\overline{\delta_{a,b}}$, see \fer{delta}. Note that $T$ is a real symmetric matrix, $[T]_{a,b}<0$ for $a\neq b$, and $[T]_{a,a}=-\sum_{b\neq a}[T]_{a,b}$. These properties imply that for $x=(x_1,\ldots,x_N)\in {\mathbb C}^N$, $\scalprod{x}{T x} = \sum_{a,b=1}^N | [T]_{a,b}|\, |x_a-x_b|^2\geq 0$. Therefore, if $[T]_{a,b}\neq 0$ for all $a\neq b$, then zero is a simple eigenvalue of $T$, with eigenvector proportional to $(1,\ldots,1)$ and all other eigenvalues of $T$ are strictly positive.

This completes the proof of Theorem \ref{thm4}. \hfill $\square$

\subsection{Proof of Theorem \ref{dynthm}}

The proof of these two theorems is based on the resolvent representation, Theorem \ref{thm2}, and on the spectral data given in Theorem \ref{thm4}. The procedure follows \cite{JP,BFS,MMS} (for the path integration deformation argument) and \cite[Theorem 3.1]{MSB} (for the reduced dynamics).

Let $\Psi_0\in\cH$ (initial state). Given $\epsilon>0$, we can find a vector $\Psi_\epsilon$ such that (a) $|\scalprod{\Psi_0}{A\Psi_0}-\scalprod{\Psi_\epsilon}{A\Omega}|<\|A\| \epsilon$, for all $A\in\fm$, where $\Omega$ is the reference state \fer{m20}, and (b) $\Psi_\epsilon$ is $U_\theta$-analytic and $U_{\bar\theta}\Psi_\eta\equiv(\Psi_\eta)_{\bar\theta}$ is in the domain of $\e^{|L_\r|/2}$. To produce $\Psi_\epsilon$, one may first find $B\in\fm'$ (commutant of $\fm$) s.t. $\|\Psi_0-B\Omega\|<\epsilon/2$ (this can be done by the cyclicity of $\Omega$) and set $\Psi_{1,\epsilon}=B^*B\Omega$. Then (a) is verified. Next, one regularizes this vector to satisfy (b), e.g. by forming  $\Psi_{2,\epsilon} =\e^{-\eta L^2_\r}\e^{-\eta D^2} \e^{-4 \eta\theta_0^2 N^2}\Psi_{1,\epsilon}$, where $D=\d\Gamma(-\i\partial_u)$ is the generator of spectral deformation and $N=\d\Gamma(\bbbone)$ is the number operator. Taking $\eta>0$ small enough gives $\Psi_\epsilon$ satisfying (a) and (b). The set of translation-analytic functionals
\begin{equation}
\label{anstates}
{\cal S}_0 = \{\scalprod{\Psi}{\,\cdot\,\Omega} : \Psi \mbox{\ satisfies (a) and (b)}\}
\end{equation}
is hence dense in the set of all states on $\fm$. The translation-analytic observables are defined by
\begin{equation} 
\label{anobs}
\fm_0 = \{ A\in\fm : A\Omega \mbox{\ is $U_\theta$-analytic}\}.
\end{equation}
Let $\omega_0\in{\cal S}_0$ and $A\in\fm_0$. Theorem \ref{thm2} gives
$$
\omega_0(\alpha^t_{\sigma,\lambda}(A)) = 
\frac{-1}{2\pi\i}\int_{\rx-\i} \e^{\i
tz}\scalprod{\Psi_{\overline\theta}}{(K_\theta(\sigma,\lambda)-z)^{-1}
(A\Omega)_\theta}\d z. 
\label{mm2'}
$$
We deform the contour of integration into the upper half-plane, as in \cite{JP,BFS,MMS}, to pick up the contributions of the poles at the resonance energies of the resolvent by means of the residue theorem.  The integral over the path $\rx-\i$ equals the integral over the path $\rx+\frac34\i{\rm Im}\theta$ plus the sum of the integrals around circles $\Gamma_{a,b}$, each enclosing exactly one eigenvalue $\varepsilon_{a,b}$ of $K_\theta(\sigma,\lambda)$. While the integral over  $\rx+\frac34\i{\rm Im}\theta$ is $O(\e^{-\frac34 t\,{\rm Im}\theta})$, the integral around a given eigenvalue $\varepsilon_{a,b}$ is
\begin{equation}
\label{new1}
\frac{-1}{2\pi \i}\oint_{\Gamma_{a,b}} \e^{\i tz}\scalprod{\Psi_{\bar\theta}}{(K_\theta(\sigma,\lambda)-z)^{-1} A \Omega}\d z = \e^{\i t\varepsilon_{a,b}(\sigma,\lambda)} \scalprod{\Psi_{\bar\theta}}{{\widetilde Q}_{a,b}(A\Omega)_\theta},
\end{equation}
where ${\widetilde Q}_{a,b} = \frac{-1}{2\pi\i}\oint_{\Gamma_{a,b}} (K_\theta(\sigma,\lambda)-z)^{-1}\d z$ is the Riesz spectral projection.

The KMS state of the uncoupled system ($\lambda=0$) is given by the standard vector $\Omega_0=\Omega_{\s,\beta}\otimes\Omega_\r$. Here, $\Omega_{\s,\beta}$ is the unique vector in the standard natural cone, the closure of $\{AJ_\s A\Omega_\s: A\in\fm_\s\}$ (recall the definition of $\Omega_\s$ and $J_\s$ given in and after \fer{m-1}), representing the system Gibbs equilibrium state (which is determined by the density matrix $\propto\e^{-\beta\sigma H_\s}$). Perturbation theory of KMS states (see \cite{BR,DJP,BFS}) tells us that 
\begin{equation}
\label{coupledkms}
\Omega_{\s\r}=\e^{-\beta L(\sigma,\lambda)/2}\Omega_0/\|\e^{-\beta L(\sigma,\lambda)/2}\Omega_0\|,
\end{equation}
where $L(\sigma,\lambda)$ is given in \fer{m11'}, is the KMS state for the interacting system. 

Consider $\sigma>0$. Since ${\rm Im}\varepsilon_{a,b}>0$ for all $a\neq b$ and ${\rm Im}\varepsilon_{a,a}>0$ for $a=2,\ldots,N$ and since $\Omega_{\s\r}$ is an invariant state, it follows by taking $t\rightarrow\infty$ that the quantity \fer{new1} for $a=b=1$ is $\scalprod{\Omega_{\s\r}}{A\Omega_{\s\r}}=\omega_{\beta,\sigma,\lambda}(A)$. The remaining contributions to the right side of \fer{1} come from the resonances bifurcating out of the origin (first sum) and those bifurcating out of $\varepsilon_{a,b}(0,\lambda)$, as $\sigma$ becomes nonzero. We have $\chi_a(A)= \langle{\Psi_{\bar\theta}},{{\widetilde Q}_{a,a}(A\Omega)_\theta}\rangle$, and a similar definition for $\chi_{a,b}$.

Consider $\sigma=0$. Then $\varepsilon_{a,a}(0,\lambda)=0$ for all $a=1,\ldots,N$. The first two terms on the right side of \fer{1} arise from the projection onto the kernel of $K_\theta(0,\lambda)$. This defines the $\chi_a$ for $\sigma=0$. The $\chi_{a,b}$ are again given by the the scalar products on the right side of \fer{new1}.

Note that the $\chi_a$ are not continuous as $\sigma\rightarrow 0$, as only the total group projection associated to the eigenvalues bifurcating out of the origin is continuous (actually analytic), but not the individual projections.

\subsection{Proof of Theorem \ref{thmreddynfinal}}

\begin{thm}[Reduced dynamics]
\label{reduceddyn}
Let $\chi_1$ be an arbitrary normalized vector in $\cH_\s$ and let $A\in\fm_\s$ be a system observable. Then we have 
\begin{eqnarray}
\lefteqn{
\scalprod{\chi_1\otimes\Omega_\r}{\e^{\i tL(\sigma,\lambda)} A \e^{-\i tL(\sigma,\lambda)}\Omega}}\nonumber\\
&=&\sum_{a,b=1}^N\e^{\i t\varepsilon_{a,b}(\sigma,\lambda)}\scalprod{\chi_1}{Q_{a,b}A\Omega_\s} \big( 1+O_\lambda(\sigma)+O(\lambda)\big) +O\big(\lambda^2\e^{-\frac34 t\theta_0}\big),
\label{l1}
\end{eqnarray}
where the $\varepsilon_{a,b}(\sigma,\lambda)$ are given in \fer{l7}. Here, 
\begin{equation}
Q_{a,b} = 
\left\{
\begin{array}{ll}
|\varphi_{a,b}\rangle\langle\varphi_{a,b}| & \mbox{if $a\neq b$}\\
|\varphi^T_a\rangle\langle\varphi^T_a| & \mbox{if $a=b$,}
\end{array}
\right.
\label{l3}
\end{equation}
where $\{\varphi_a^T\}_{a=1}^N$ is the orthonormal basis of eigenvectors of $T$, \fer{mmm6}, so that $T\varphi_a^T=\xi_a\varphi_a^T$.
\end{thm}

\noindent
{\bf Proof of Theorem \ref{reduceddyn}.\ } Take the representation \fer{mm2} for a fixed $\theta$. The integral over the path $\rx-\i$ equals the integral over the path $\rx+\frac34\i{\rm Im}\theta$ plus the sum of the integrals around circles $\Gamma_{a,b}$, each enclosing exactly one eigenvalue $\varepsilon_{a,b}$ of $K_\theta(\sigma,\lambda)$. While the integral over  $\rx+\frac34\i{\rm Im}\theta$ is $O(\lambda^2\e^{-\frac34 t\,{\rm Im}\theta})$ (see Proposition \ref{prop7}), the integral around a given eigenvalue $\varepsilon_{a,b}$ is
$$
\frac{-1}{2\pi \i}\oint_{\Gamma_{a,b}} \e^{\i tz}\scalprod{\chi_1\otimes\Omega_\r}{(K_\theta(\sigma,\lambda)-z)^{-1} A \Omega}\d z = \e^{\i t\varepsilon_{a,b}} \scalprod{\chi_1\otimes\Omega_\r}{{\widetilde Q}_{a,b}A\Omega},
$$
where ${\widetilde Q}_{a,b} = \frac{-1}{2\pi\i}\oint_{\Gamma_{a,b}} (K_\theta(\sigma,\lambda)-z)^{-1}\d z$ is the Riesz spectral projection. By perturbation theory, we have, for $a\neq b$,
$$
\widetilde{Q}_{a,b}=|\varphi_{a,b}\rangle\langle\varphi_{a,b}|\otimes|X_{a,b}\rangle\langle X_{a,b}^*| +O_\lambda(\sigma)=|\varphi_{a,b}\rangle\langle\varphi_{a,b}|\otimes|\Omega_\r\rangle\langle \Omega_\r| +O_\lambda(\sigma)+O(\lambda).
$$
Similarly, we have $\widetilde{Q}_{a,a}=|\varphi_a^T\rangle\langle\varphi^T_a|\otimes|\Omega_\r\rangle\langle\Omega_\r|+O_\lambda(\sigma)$. (Note that $T$ is self-adjoint.) This completes the proof of Theorem \ref{reduceddyn}.\hfill $\square$

We now prove Theorem \ref{thmreddynfinal}. Let $\rho_0$ be the initial density matrix of the small system. It is represented by a normalized vector $\chi$ in the GNS space $\cH_\s$. By the cyclicity of $\Omega_\s$ there is a unique element $B'$ in the commutant $\fm_{\cx^N}'=\bbbone_{\cx^N}\otimes{\cal B}(\cH_\s)$ such that $\chi=B'\Omega_\s$. The evolution of the reduced density matrix elements $[\rho_t]_{a,b}=\scalprod{\varphi_a}{\rho_t\varphi_b}$ is given by
\begin{eqnarray}
{}[\rho_t]_{a,b} &=& \scalprod{\chi\otimes\Omega_\r}{\e^{\i t L(\sigma,\lambda)} (|\varphi_b\rangle\langle\varphi_a|\otimes\bbbone_{\cx^N})\e^{-\i t L(\sigma,\lambda)}\chi\otimes\Omega_\r}\nonumber\\
&=& \scalprod{\chi\otimes\Omega_\r}{B'\e^{\i t L(\sigma,\lambda)}( |\varphi_b\rangle\langle\varphi_a|\otimes\bbbone_{\cx^N})\e^{-\i t L(\sigma,\lambda)}\Omega}.
\end{eqnarray}
We can thus use Theorem \ref{reduceddyn}. The main term on the right side of \fer{l1} is
\begin{equation}
\sum_{c,d=1}^N\e^{\i t\varepsilon_{c,d}(\sigma,\lambda)}\scalprod{\chi}{B'Q_{c,d} ( |\varphi_b\rangle\langle\varphi_a|\otimes\bbbone_{\cx^N}) \Omega_\s}=\frac{1}{\sqrt N} \sum_{c,d=1}^N\e^{\i t\varepsilon_{c,d}(\sigma,\lambda)}\scalprod{\chi}{B'Q_{c,d} \, \varphi_{b,a}},
\label{l2}
\end{equation}
by the definition \fer{m-1} of $\Omega_\s$. If $a\neq b$ then, according to \fer{l3}, $Q_{c,d} \varphi_{b,a}$ vanishes, except when $c=b$ and $d=a$, in which case it equals $\varphi_{b,a}$. Then we have $\scalprod{\chi}{B'\varphi_{b,a}}=\sqrt N \scalprod{\chi}{B'(|\varphi_b\rangle\langle\varphi_a|\otimes\bbbone_{\cx^N}|)\Omega_\s}=\sqrt N\,  [\rho_0]_{a,b}$. We conclude that for $a\neq b$, the main term of $[\rho_t]_{a,b}$ is $\e^{\i t\varepsilon_{b,a}(\sigma,\lambda)}[\rho_0]_{a,b}$. This shows \fer{l5}. Relation \fer{l9} is proven in the same way. \hfill $\square$

\subsection{Using the Feshbach map}

Zero is an eigenvalue of $K_\theta(0,0)$ of multiplicity $N^2$. By a simple Riesz projection argument, one shows that, for $\sigma$ and $\lambda$ small, $K_\theta(\sigma,\lambda)$ has $N^2$ eigenvalues in the vicinity of the origin. The size of the eigenvalues can be estimated as follows. Suppose that $z\neq 0$ and ${\rm Im}z<\frac12{\rm Im}\theta$, so that $z$ is in the resolvent set of $K_\theta(0,0)$. If the series
\begin{equation}
( K_\theta(0,0)-z)^{-1}\sum_{n\geq 0} \big[(\sigma  L_\s +\lambda  I_\theta)( K_\theta(0,0)-z)^{-1}\big]^n
\label{fesh0}
\end{equation}
converges, then $z$ belongs to the resolvent set of $K_\theta(\sigma,\lambda)$ and \fer{fesh0} equals $(K_\theta(\sigma,\lambda)-z)^{-1}$. Therefore, if $z$ is a (non-zero) eigenvalue of $K_\theta(\sigma,\lambda)$, then we must have
\begin{equation}
\|(\sigma L_\s +\lambda I_\theta)(K_\theta(0,0)-z)^{-1}\|\geq 1.
\label{fesh0.1}
\end{equation}
Using standard bounds on the interaction, we see that \fer{fesh0.1} implies that there are constants $C,c>0$ s.t. if $\sigma,|\lambda|<c$, then 
\begin{equation}
|z| < C(\sigma+|\lambda|).
\label{fesh0.2}
\end{equation}
Estimate \fer{fesh0.2} is a bound on the eigenvalues of $K_\theta(\sigma,\lambda)$ in the vicinity of the origin. The eigenvalues can be tracked using the Feshbach map. Namely, $z\in\mathbb C$, ${\rm Im}z<\frac12{\rm Im}\theta$ is an eigenvalue of $K_\theta(\sigma,\lambda)$ if and only if it is an eigenvalue of the operator
\begin{equation}
{\cal F}_z = P_\r\left( \sigma L_\s -\lambda^2 I_\theta( K_\theta(\sigma,\lambda)-z)^{-1}I_\theta\right) P_\r
\label{fesh1}
\end{equation}
which acts on the smaller space ${\rm Ran}P_\r={\mathbb C}^N\otimes{\mathbb C}^N$. Recall that $P_\r=|\Omega_\r\rangle\langle\Omega_\r|$. By expanding the resolvent around $z=0$, $\sigma=0$ and $\lambda=0$, taking into account \fer{fesh0.2}, we have
\begin{equation}
{\cal F}_z= P_\r\left( \sigma L_\s -\lambda^2I_\theta K_\theta(0,0)^{-1}I_\theta\right) P_\r +O\Big(\lambda^2\big(\sigma+|\lambda|\big)\Big),
\label{fesh2}
\end{equation}
provided $z$ is an eigenvalue of $K_\theta(\sigma,\lambda)$ and $\sigma, |\lambda|<c$.  An elementary calculation shows that the operator ${\cal F}_z$, viewed as acting on ${\rm Ran}P_\r$, has the form
\begin{equation}
{\cal F}_z=\sigma L_\s  -\lambda^2\big(\alpha G^2\otimes \bbbone -\alpha G\otimes  G+\overline{\alpha}G\otimes  G -\overline{\alpha}\bbbone\otimes  G^2 \big)+ O\Big(\lambda^2\big(\sigma+|\lambda|\big)\Big),
\label{fesh3}
\end{equation}
where ${\cal C}$ is defined after \fer{m11} and $\alpha= \frac12 \scalprod{g}{|k|^{-1}g}- \frac{\i}{2}\pi\xi(0)$, with $\xi(0)$ given in \fer{xinot}. Note that the quadratic term in $\lambda$ is diagonal in the basis $\varphi_{a,b}$, 
\begin{eqnarray}
\lefteqn{
 -\lambda^2\big(\alpha G^2\otimes \bbbone -\alpha G\otimes  G +\overline{\alpha}G\otimes  G -\overline{\alpha}\bbbone\otimes  G^2 \big)}\nonumber\\
&& =  -\frac{\lambda^2}{2}\sum_{a,b=1}^N \big(\scalprod{g}{|k|^{-1}g}(g^2_a-g^2_b) -\i\pi\xi(0)(g_a-g_b)^2\big) |\varphi_{a,b}\rangle\langle\varphi_{a,b}|.
\label{fesh4}
\end{eqnarray}
We conclude from the isospectrality of the Feshbach map and \fer{fesh3}, \fer{fesh4} that the eigenvalues of $K_\theta(0,\lambda)$ are given by
$ -\frac{\lambda^2}{2}\big(\scalprod{g}{|k|^{-1}g}(g^2_a-g^2_b) -\i\pi\xi(0)(g_a-g_b)^2\big)$,
modulo a remainder $O(\lambda^2(\sigma+|\lambda|))$. This is compatible with the result of Theorem \ref{thm3}. However, from that Theorem, we know in addition that the remainder actually vanishes.

\subsection{The spin-boson system}
\label{sectspinboson}

The Feshbach operator \fer{fesh3} is represented in the energy basis $\{\phi_{+,+},\ \phi_{+,-},\ \phi_{-,+}\ \phi_{-,-}\}$, where $\phi_{+,-}=\phi_+\otimes\phi_-$ (etc) and $S^z\phi_\pm=\pm\frac12\phi_\pm$, by the matrix
\begin{eqnarray}
{\cal F}_z&=& W + O\Big(\lambda^2\big(\sigma+|\lambda|\big)\Big),\\
W&=& \left(
\begin{array}{cccc}
\i\frac{\lambda^2}{4}\pi\xi(0) & 0 & 0  & -\i\frac{\lambda^2}{4}\pi \xi(0)\\
0 & \sigma+\i\frac{\lambda^2}{4}\pi \xi(0)& -\i\frac{\lambda^2}{4}\pi \xi(0) & 0 \\
0 & -\i\frac{\lambda^2}{4}\pi \xi(0) & -\sigma+\i\frac{\lambda^2}{4}\pi \xi(0) & 0 \\
-\i\frac{\lambda^2}{4}\pi \xi(0)& 0 &0  & \i\frac{\lambda^2}{4}\pi
\xi(0)
\end{array}
\right).
\end{eqnarray}
The four eigenvalues of $W$ are
\begin{equation}
w_1=0,\quad w_2=\i \frac{\lambda^2}{2}\pi \xi(0), \quad w_{3,4}=\i\frac{\lambda^2}{4}\pi\xi(0)\pm\i \sqrt{\frac{\lambda^4}{16}\pi^2\xi(0)^2-\sigma^2},
\label{evals}
\end{equation}
where the square root is the principal branch with branch cut on the negative real axis. The corresponding eigenvectors of $W$ are
\begin{equation}
\chi_1=\frac{1}{\sqrt 2}\left[\begin{array}{c} 1\\
0\\0\\1\end{array}\right],\ \ 
\chi_2=\frac{1}{\sqrt 2}\left[\begin{array}{c} 1\\
0\\0\\-1\end{array}\right],\ \ 
\chi_3= \frac{1}{1+r^2}\left[\begin{array}{c} 0\\
1\\
r\\0
\end{array}\right],\ \ 
\chi_4=\frac{1}{1+r^2}\left[\begin{array}{c} 0\\
-r\\1\\0\end{array}\right],
\label{evect}
\end{equation}
where $r=\frac{-4\i \gamma -\sqrt{\pi^2\xi(0)^2 -16\gamma^2}}{\pi\xi(0)}$ with $\gamma=\frac{\sigma}{\lambda^2}$. 
The eigenvalues of the adjoint $W^*$ are the complex conjugates $\overline w_j$ and the corresponding eigenvectors are 
\begin{equation}
\chi_1^*=\frac{1}{\sqrt 2}\left[\begin{array}{c} 1\\
0\\0\\1\end{array}\right],\ \ 
\chi_2^*=\frac{1}{\sqrt 2}\left[\begin{array}{c} 1\\
0\\0\\-1\end{array}\right],\ \
\chi_3^*=\left[\begin{array}{c} 0\\
1\\
\overline r \\0\end{array}\right],\ \ 
\chi_4^*=\left[\begin{array}{c} 0\\
-\overline r\\1\\0\end{array}\right].
\label{evect*}
\end{equation}
The eigenvectors are normalized as $\scalprod{\chi_i}{\chi_i^*}=1$ and $\langle\chi_i, \chi_j^*\rangle=0$ if $i\neq j$. The reduced spin density matrix, represented in the energy basis $\phi_\pm$, is given by (proceed as for Theorem \ref{thmreddynfinal} or see \cite[Theorem 2.1]{mjmp} and \cite{MSB})
\begin{equation}
[\rho_t]^z_{m,n}\doteq\sum_{j=1}^4\e^{\i t w_j}\sum_{k,l=\pm}[\rho_0]^z_{l,k}\scalprod{\phi_{k,l}}{\chi_j}\scalprod{
\chi_j^*}{\phi_{n,m}}.
\label{u1}
\end{equation}
Here, we take $m,n,k,l$ to stand for either $+$ or $-$, and $\doteq$ means that we approximate the true resonances $\varepsilon$ (the eigenvalues of ${\cal F}_z$) by the $w$ and we neglect additive $O(\lambda^2)$ terms (uniform in $t\geq 0$) on both sides. Using the explicit formulas \fer{evect}, \fer{evect*} for the eigenvectors $\chi_j$, $\chi^*_j$, we arrive at
\begin{equation}\label{1.106}
\begin{split}
[\rho_t]^z_{+,+}\doteq&\textstyle\frac{1}{2}+\frac{1}{2}\e^{\i t
w_2}([\rho_0]^z_{+,+}-[\rho_0]^z_{-,-}),\\
[\rho_t]^z_{+,-}\doteq &\textstyle \frac{r}{r^2+1}{\e^{\i
tw_3}(r[\rho_0]^z_{+,-}+[\rho_0]^z_{-,+})+\frac{1}{r^2+1}\e^{\i
tw_4}([\rho_0]^z_{+,-}-r[\rho_0]^z_{-,+})}.
\end{split}
\end{equation}

\appendix


\renewcommand{\theresultcounter}{\Alph{section}.\arabic{resultcounter}}




\section{Invariant states}
\label{appb}

{\bf Invariant system-reservoir states.\ } Let $L_{\rm standard}=L_0(\sigma)+\lambda V-\lambda J VJ$ be the  {\em standard Liouvillian} and let ${\cal P}$ be the closure of the set $\{AJA\Omega\, :\, A\in\fm\}$ (the natural positive cone associated to $(\fm,\Omega)$; see also \fer{m20}). There is a one-to-one correspondence between normalized vectors in ${\rm Ker} L_{\rm standard}\cap\cal P$ and normal states on $\fm$ which are invariant under the dynamics generated by $L$, \fer{m11'} (see for instance \cite{DJP}). 

For $\sigma=0$, the standard Liouvillian has a direct sum decomposition as in \fer{kab}, with `blocks' $L_{{\rm standard},a,b}=L_\r+\lambda \{g_a\Phi(g_\beta)-g_bJ \Phi(g_\beta)J\}$. One can perform the spectral analysis of this operator in the same way as we do for $K(0,\lambda)$ to see that ${\rm Ker}L_{\rm standard}= {\rm span}\{\varphi_a\otimes\varphi_a\otimes\Omega_{\r,a}\}_{a=1}^N$, where 
\begin{equation}
\label{mexpl}
\Omega_{\r,a}=\frac{\e^{-\beta(L_\r+\lambda g_a\Phi(g_\beta))/2}\Omega_\r}{\|\e^{-\beta(L_\r+g_a\Phi(g_\beta))/2}\Omega_\r\|}
\end{equation}
is the reservoir KMS state with respect to the dynamics generated by the Liouvillian $L_\r+\lambda g_a\Phi(g_\beta)$, denoted by $\omega_{\r,a}$. This `perturbed' KMS state belongs to the standard natural cone associated to $(\fm_\r,\Omega_\r)$ (see e.g. \cite{DJP}) and hence $\varphi_a\otimes\varphi_a\otimes\Omega_{\r,a}\in\cal P$.


For $\sigma>0$ and under the condition that $K_\theta(\sigma,\lambda)$ has one-dimensional kernel, the only invariant state is the coupled equilibrium $\Omega_{\s\r}$ introduced in \fer{coupledkms}.

{\bf Invariant initial states of the small system for $\sigma=0$.\ } The explicit expression \fer{m31} shows that ${\cal M}_{0,\lambda}$, the manifold of invariant initial system states, is the set of density matrices which are diagonal in the eigenbasis $\{\varphi_a\}_{a=1}^N$ of $G$. Let $\rho_0$ be a given initial density matrix of the small system and set $\tau=\sum_a [\rho_0]_{a,a}|\varphi_a\rangle\langle\varphi_a|$. Then ${\rm dist}({\cal M}_{0,\lambda},\rho_0)=\|\tau-\rho_0\|_1$. To see this, let $\tau_n$ be a sequence in ${\cal M}_{0,\lambda}$ such that $\lim_{n\rightarrow\infty} \|\tau_n-\rho_0\|_1={\rm dist}({\cal M}_{0,\lambda},\rho_0)$. By the equivalence of the trace norm and the norm $\|\rho\|_{\rm max}= \max_{a,b}|\scalprod{\varphi_a}{\rho\varphi_b}|\equiv\max_{a,b}|[\rho]_{a,b}| $, we have 
$$
 \|\tau_n-\rho_0\|_1\geq c \|\tau_n-\rho_0\|_{\rm max}\geq c \max_a \big| [\tau_n]_{a,a}-[\rho_0]_{a,a}\big|,
$$
for some constant $c>0$. It follows that $\lim_{n\rightarrow\infty} \max_a | [\tau_n]_{a,a}-[\rho_0]_{a,a}|=0$ and therefore $\lim_{n\rightarrow\infty} \|\tau_n-\tau\|_1=0$. This shows that ${\rm dist}({\cal M}_{0,\lambda},\rho_0)=\|\tau-\rho_0\|_1$. As the dynamics leaves the diagonal invariant, we also have ${\rm dist}({\cal M}_{0,\lambda},T_{0,\lambda}(t)\rho_0)=\|\tau-T_{0,\lambda}(t)\rho_0\|_1$.  Again by the equivalence of norms, there is a $C>0$ s.t. 
$$
\|\tau-T_{0,\lambda}(t)\rho_0\|_1\leq C\max_{a,b: a\neq b} | [T_{0,\lambda}(t)\rho_0]_{a,b}|\leq C \e^{-\lambda^2\gamma_G\Gamma(t)}\max_{a,b: a\neq b} | [\rho_0]_{a,b}|,
$$
where we use \fer{m31} in the last inequality. Finally, $\max_{a,b: a\neq b} | [\rho_0]_{a,b}|\leq c \|\tau-\rho_0\|_1$. The statement about orbital stability after \fer{*} follows. The asymptotic linearity of $\Gamma(t)$ follows from \fer{m32}. In three dimensions, $\lim_{t\rightarrow\infty}\Gamma(t)=\infty$ if the infra-red behaviour of the coupling form factor is $g(k)\sim |k|^{-1/2}$ as $k\sim 0$, see \fer{delta}. See also \cite{PSE}.

\medskip

{\bf Absence of invariant initial system states for $\sigma>0$.} Suppose that zero is a simple eigenvalue of $K_\theta(\sigma,\lambda)$. Then for $\sigma>0$, the set of invariant initial system states ${\cal M}_{\sigma,\lambda}$ is empty. Indeed, by the property of return to equilibrium, $\lim_{t\rightarrow\infty} T_{\sigma,\lambda}(t)\rho_0 = \rho_*$ for all initial states $\rho_0$, where $\rho_*$ is the reduction to the small system of the coupled system-reservoir KMS state $\Omega_{\s\r}$ (see \fer{coupledkms}). Therefore, $\rho_*$ is the only possible element in ${\cal M}_{\sigma,\lambda}$. However, that $\rho_*\not\in{\cal M}_{\sigma,\lambda}$ can be seen as follows. For any $A\in{\cal B}(\cx^N)$ we have
\begin{equation*}
\frac{\d}{\d t}\Big|_{t=0} {\rm Tr}_{\cx^N}(T_{\sigma,\lambda}(t)\rho_* \,A) =\scalprod{\Omega_*\otimes\Omega_\r}{ \i [L(\sigma,\lambda), A\otimes\bbbone_\s\otimes\bbbone_\r]\Omega_*\otimes\Omega_\r},
\end{equation*}
where $\Omega_*$ is the vector representative of $\rho_*$. The commutator in the last expression equals $\sigma[H_\s,A]\otimes\bbbone_\s\otimes\bbbone_\r + \lambda[G,A]\otimes\bbbone_\s\otimes\Phi(g_\beta)$. Therefore, the above derivative is zero if and only if $\scalprod{\Omega_*}{([H_\s,A]\otimes\bbbone_\s)\Omega_*}=\scalprod{\Omega_{\s\r}}{([H_\s,A]\otimes\bbbone_\s\otimes\bbbone_\r)\Omega_{\s\r}}=0$. By expanding $\Omega_{\s\r}\propto \Omega_0 -\frac{\lambda}{2}\int_0^\beta\e^{-s L_0/2} V\Omega_0+O(\lambda^2)$ (see \fer{coupledkms}), we obtain
\begin{equation}
\scalprod{\Omega_{\s\r}}{([H_\s,A]\otimes\bbbone_\s\otimes\bbbone_\r)\Omega_{\s\r}}=\frac{\lambda^2\sigma }{2}\sum_{k,l=1}^N (E_k-E_l) \langle GP_kAP_l G\rangle_{\s,\beta} \, f_{k,l} +O(\lambda^4),
\label{rs}
\end{equation}
where $P_k$ is the spectral projection associated to the eigenvalue $E_k$ of $H_\s$, the average $\langle\cdot\rangle_{\s,\beta}$ is taken in the state $\Omega_{\s,\beta}$ and where $f_{k,l}= \int_{\rx\times S^2}|g_\beta(u,\vartheta)|^2\,\frac{(\e^{\beta u/2}-1)(\e^{-\beta u/2}-1)}{u^2}\d u\d\vartheta +O(\sigma)$. For small $\sigma$, we have $f_{k,l}<0$ for all $k,l$. By choosing an $A$ s.t. the right side of \fer{rs} does not vanish we obtain $\frac{\d}{\d t}|_{t=0} {\rm Tr}_{\cx^N}(T_{\sigma,\lambda}(t)\rho_* \,A) \neq 0$, so $\rho_*$ is not invariant.

\section{Proof of Theorem \ref{thm2}}
\label{appa}
Throughout the proof, we do not write the dependence of operators on $(\sigma,\lambda)$ (i.e., we write $L$ for $L(\sigma,\lambda)$, and so on).

Let $s\in\cx$, $|s|<1/2+\epsilon$, where $\epsilon$ is the constant in Assumption A2. Using the expression $\Delta=\bbbone_{\cH_\s}\otimes\e^{-\beta L_\r}$ for the modular operator, we get
\begin{eqnarray}
\Delta^{\i\overline s}V\Delta^{-\i\overline s} &=& G\otimes\bbbone_{\cx^N}\otimes\e^{-\i\beta\overline s L_\r}\Phi(g_\beta) \e^{\i\beta\overline s L_\r}\nonumber\\
&=& G\otimes\bbbone_{\cx^N}\otimes\frac{1}{\sqrt{2}}\big( a^*(\e^{-\i\beta\overline s u}g_\beta) + a(\e^{-\i\beta su}g_\beta)\big).
\label{m30}
\end{eqnarray}
This operator is well-defined and strongly analytic in $\bar s$ on $\dom(N^{1/2})$, due to assumption (A2). On $\dom(L_0)\cap\dom(N^{1/2})$ we define the family of strongly analytic operators in $s$,
\begin{eqnarray}
K^{(s)} &=& L_0 + \lambda I^{(s)},\\
I^{(s)} &=& V -\lambda V'^{(s)},\\
V'^{(s)} &=& \Delta^{-\i s} JVJ\Delta^{\i s} = J\Delta^{\i\overline s}V\Delta^{-\i\overline s}J.
\label{mm3}
\end{eqnarray}
This family has been introduced in \cite{MMS}. It interpolates between the self-adjoint $K^{(0)}$ and the operator $K^{(-\i/2)}=K$ (see \fer{mm0}).
\begin{prop}
\label{prop3}
Let $I^{(s)}(t)=\e^{\i tL_0}I^{(s)}\e^{-\i tL_0}$ and recall the definition \fer{m20} of the reference state $\Omega$. The Dyson series
\begin{equation}
\sum_{n\geq 0}(\i\lambda)^n\int_0^t\d t_1\int_0^{t_1}\d
t_2\cdots\int_0^{t_n-1}\d t_n \ I^{(s)}(t_n)I^{(s)}(t_{n-1}) \cdots
I^{(s)}(t_1)\Omega \label{mm4}
\end{equation}
converges for all $\lambda\in\rx$ and is analytic in $s$ for $|s|<1/2+\epsilon$.
\end{prop}

{\bf Proof of Proposition \ref{prop3}.\ } Let $\psi_\nu\in {\rm Ran}\,P(N\leq\nu)$ (spectral projection of $N$ onto subspace with at most $\nu$ particles). Since the interaction operator $I^{(s)}$ changes the particle number by at most one, we have
\begin{equation}
\nonumber
\begin{split}
&I^{(s)}(t_n)I^{(s)}(t_{n-1})\cdots I^{(s)}(t_1)\psi_\nu\\
=&\e^{\i t_nL_0}I^{(s)}P(N\leq\nu+n-1)\e^{-\i t_nL_0}\cdots \e^{\i
t_1L_0}I^{(s)}P(N\leq\nu)\e^{-\i t_1L_0}\psi_\nu.
\end{split}
\end{equation}
The standard bounds $ \| a^*(f)(N+1)^{-1/2}\|\leq \|f\|$ and $\|a(f)(N+1)^{-1/2}\|\leq \|f\|$ give $\| I^{(s)}(N+1)^{-1/2}\| \leq 4M$, where $M:=(\int|e^{(\frac{1}{2}+\epsilon)\beta|u|}g_\beta(u,\sigma)|^2\d
u\d \sigma)^{\frac{1}{2}}<\infty$ due to assumption (A2). Hence
\begin{equation}
\|I^{(s)}(t_n)I^{(s)}(t_{n-1})\cdots
I^{(s)}(t_1)\psi_\nu\| \leq\sqrt{(\nu+1)\cdots(\nu+n)} (4M)^n \|\psi_\nu\|,
\label{p1.3}
\end{equation}
uniformly in $s$. This and the
analyticity of $I^{(s)}(t_n)I^{(s)}(t_{n-1})\cdots
I^{(s)}(t_1)\psi_\nu$ imply that \eqref{mm4} is analytic in $s$ for 
$|s|<\frac{1}{2}+\epsilon$. This proves Proposition \ref{prop3}.\qed

\bigskip

We define an operator denoted $\e^{\i tK^{(s)}}$, on the dense set ${\frak M}\Omega$, by
\begin{equation}
\e^{\i tK^{(s)}}\Omega :=\fer{mm4} \mbox{\quad and\quad} \e^{\i t
K^{(s)}}A\Omega:=\e^{\i tL}A\e^{-\i t L}\e^{\i tK^{(s)}}\Omega
\label{mm7}
\end{equation}
for $A\in \frak M$.

\begin{prop}
\label{prop4}
We have $\e^{\i tK^{(-\i/2)}} A\Omega = \e^{\i tL}A\e^{-\i tL}\Omega$, for all $A\in\frak M$.
\end{prop}

{\bf Proof of Proposition \ref{prop4}.\ } It suffices to show that
$\e^{\i tK^{(-\i/2)}}\Omega=\Omega$. Note that $(G\otimes
\bbbone)\Omega_\s=(\bbbone \otimes {\cal C}G{\cal C})\Omega_\s$ (see after \fer{m11} for the definition of $\cal C$), $J\Delta^{\frac{1}{2}}\Omega_\r=\Omega_\r$ and that $\Phi(g_\beta)$ is
selfadjoint. Thus,
\begin{equation}
\nonumber
\begin{split}
I^{(-\i/2)}\Omega=&[G\otimes \bbbone\otimes
\Phi(g_\beta)-\bbbone\otimes G\otimes
J\Delta^{\frac{1}{2}}\Phi(g_\beta)J\Delta^{\frac{1}{2}}]\Omega_\s\otimes\Omega_\r\\
=&(G\otimes
\bbbone)\Omega_\s\otimes[\Phi(g_\beta)\Omega_\r-J\Delta^{\frac{1}{2}}\Phi(g_\beta)J\Delta^{\frac{1}{2}}\Omega_\r]=0.
\end{split}
\end{equation}
It now follows directly from  \fer{mm7} and \eqref{mm4} that $\e^{\i tK^{(-\i/2)}}\Omega=\Omega$. \qed

\bigskip

Let $\psi = A\Omega$. Since $K^{(s)}$ is self-adjoint for $s\in\rx$, we have
\begin{equation}
\scalprod{\phi}{\e^{\i tK^{(s)}}\psi} = \frac{-1}{2\pi\i}\int_{\rx-\i} \e^{\i tz} \scalprod{\phi}{(K^{(s)}-z)^{-1}\psi}\d z,\quad s\in\rx.
\label{mm5}
\end{equation}
Next we perform the spectral deformation. By analyticity the scalar product in the integrand of \fer{mm5} equals
 $\scalprod{\phi_{\overline\theta}}{(K_\theta^{(s)}-z)^{-1}\psi_\theta}$, for all $|\theta|<\theta_0$. Here, $K^{(s)}_\theta=L_{0,\theta}+\lambda I^{(s)}_\theta$ is the analytic extension of $U_\theta K^{(s)}U^*_\theta$ to complex $|\theta|<\theta_0$. Thus we obtain
\begin{equation}
\scalprod{\phi}{\e^{\i tK^{(s)}}\psi}
= \frac{-1}{2\pi\i}\int_{\rx-\i}\e^{\i tz} \scalprod{\phi_{\overline\theta}}{(K_\theta^{(s)}-z)^{-1}\psi_\theta} \d z,\quad s\in\rx.
\label{mm6}
\end{equation}
From now on we take $\theta$ to be a fixed $\i\theta$, for some $0<\theta<\theta_0$.
\begin{prop}
\label{prop5}
Both sides in \fer{mm6} have an analytic extension to $s\in\cx$, $|s|<1/2+\epsilon$. Since they are equal for real $s$ we have (by the identity principle) that \fer{mm6} stays valid for all $|s|<1/2+\epsilon$.
\end{prop}

Taking the value $s=-\i/2$ in \fer{mm6}, together with Proposition \ref{prop4}, gives relation \fer{mm2} and hence proves Theorem \ref{thm2}.

{\bf Proof of Proposition \ref{prop5}.\ } Analyticity of the l.h.s.
of \fer{mm6} is immediate from Proposition \ref{prop3} and relations
\fer{mm7}. To prove the analyticity of r.h.s. of \fer{mm6}, we first
prove the convergence of the improper Riemann integral. The second resolvent equation gives
\begin{equation}
\begin{split}
(K_\theta^{(s)}-z)^{-1} =(L_{0\theta}-z)^{-1}
+(L_{0\theta}-z)^{-1}\lambda
I_\theta^{(s)}(K_\theta^{(s)}-z)^{-1}.
\label{1.18}
\end{split}
\end{equation}
Accordingly, the right side of \fer{mm6} consists of two terms. The first one, coming from the uncoupled resolvent, equals $\scalprod{\phi}{\e^{\i tL_0}\psi}$. Hence we only need to show the convergence of the integral
\begin{equation}
\frac{-1}{2\pi\i}\int_{\rx-\i}\e^{\i tz} \scalprod{\phi_{\overline\theta}}{(L_{0\theta}-z)^{-1}\lambda
I_\theta^{(s)}(K_\theta^{(s)}-z)^{-1}\psi_\theta} \d z.
\label{1.19}
\end{equation}
Consider
\begin{equation}
\begin{split}
(K_\theta^{(s)}-z)^{-1}&=(L_{0\theta}+\lambda
I_\theta^{(s)}-z)^{-1}\\
&=(L_{0\theta}-z)^{-\frac{1}{2}}[\bbbone-(L_{0\theta}-z)^{-\frac{1}{2}}\lambda
I_\theta^{(s)}(L_{0\theta}-z)^{-\frac{1}{2}}]^{-1}(L_{0\theta}-z)^{-\frac{1}{2}}.\label{1.21}
\end{split}
\end{equation}
Since $I_\theta^{(s)}(N+1)^{-\frac{1}{2}}$ is bounded and ($z=x-\i$)
\begin{equation}
\|(N+1)^{\frac{1}{2}}(L_{0\theta}-z)^{-\frac{1}{2}}\|=\sup_{n\geq
0,
l\in\rx} \frac{\sqrt{n+1}}{\sqrt[4]{(l-x)^2+(\theta n+1)^2}} \leq\frac{2}{\sqrt \theta},
\label{1.22}
\end{equation}
we have $\|(L_{0\theta}-z)^{-\frac{1}{2}}\lambda I_\theta^{(s)}(L_{0\theta}-z)^{-\frac{1}{2}}\| <1/2$, for $|\lambda|$ small enough. It follows from \fer{1.21} that
\begin{equation}
(K_\theta^{(s)}-z)^{-1}=(L_{0\theta}-z)^{-\frac{1}{2}}B(L_{0\theta}-z)^{-\frac{1}{2}}\label{1.23},
\end{equation}
where $B$ is a bounded operator satisfying $\|B\|\leq \frac{1}{1-1/2}=2$. This and \fer{1.22} imply that
\begin{equation}
\|\lambda
I_\theta^{(s)}(K_\theta^{(s)}-z)^{-1}(L_{0\theta}-z)^{\frac{1}{2}}\|\leq C |\lambda|,
\label{1.24}
\end{equation}
for some constant $C$. We estimate the integrand in \fer{1.19} as 
\begin{eqnarray}
\lefteqn{
\left|\scalprod{\phi_{\overline\theta}}{(L_{0\theta}-z)^{-1}\lambda
I_\theta^{(s)}(K_\theta^{(s)}-z)^{-1}\psi_\theta}\right|\nonumber}\\
&&
\leq  C |\lambda| \ \|(L_{0\theta}^*-\bar z)^{-1}\phi_{\bar \theta}\|\ 
\|(L_{0\theta}-z)^{-\frac{1}{2}}\psi_\theta\|\nonumber\\
&&\leq C |\lambda| \big\{(1+|x|)^{\frac{1}{2}+\eta}\|(L_{0\theta}^*-\bar
z)^{-1}\phi_{\bar
\theta}\|^2+(1+|x|)^{-\frac{1}{2}-\eta}\|(L_{0\theta}-z)^{-\frac{1}{2}}\psi_\theta\|^2\big\}\nonumber\\
&&=C |\lambda| \{ S_1(x)+S_2(x)\}.
\label{1.25}
\end{eqnarray}
The last line defines the two functions $S_1$ and $S_2$ of $x={\rm Re}z$. Here we use the inequality $ab\leq \alpha
a^2+b^2/\alpha$, for $\alpha=(1+|x|)^{1/2+\eta}$, where $0<\eta<1/2$. We have
\begin{equation}
\begin{split}
S_1(x)=&(1+|x|)^{\frac{1}{2}+\eta}\scalprod{\phi_{\bar
\theta}}{(L_{0\theta}-z)^{-1}(L_{0\theta}^*-\bar z)^{-1}\phi_{\bar
\theta}}\\
=&\sum_{n=0}^{\infty}(1+|x|)^{\frac{1}{2}+\eta}\scalprod{\phi_{\bar
\theta}}{(L_{0\theta}-z)^{-1}(L_{0\theta}^*-\bar
z)^{-1}P(N=n)\phi_{\bar
\theta}}\\
=&\sum_{n=0}^\infty
\int_{{\mathbb R}}\frac{(1+|x|)^{\frac12+\eta}}{(l-x)^2+(\theta
n+1)^2}d \mu_n(l),
\label{1.26}
\end{split}
\end{equation}
where $\d \mu_n$ is the spectral measure of $L_R$ associated to
vector $P(N=n)\phi_{\bar \theta}$ and $P(N=n)$ is the spectral projection onto the $n$ particle sector. By Fubini's theorem,
\begin{equation}
\int_{{\mathbb R}} S_1(x)\d x =\sum_{n=0}^\infty\int_{{\mathbb R}}\big[\int_{{\mathbb R}}\frac{(1+|x|)^{\frac{1}{2}+\eta}}{(l-x)^2+(\theta
n+1)^2}\d x\big]\d \mu_n(l).
\label{1.27}
\end{equation}
The integral over $x$ is bounded above by
$$
\int_{{\mathbb R}}\frac{(1+|x+l|)^{\frac{1}{2}+\eta}}{x^2+1}\d x \leq \int_{{\mathbb R}}\frac{(1+|x|)^{\frac{1}{2}+\eta}+|l|^{\frac{1}{2}+\eta}}{x^2+1}\d x \leq C_\eta+\pi |l|^{\frac12+\eta}.
$$
We use here that $(a+b)^r\leq a^r+b^r$ for $a,b\geq 0$, $0<r<1$. It follows from \fer{1.27} and this estimate that 
\begin{equation}
\int_{{\mathbb R}} S_1(x)\d x \leq \scalprod{\phi_{\bar\theta}}{(C_\eta+\pi|L_\r|^{\frac12+\eta})\phi_{\bar\theta}}<\infty.
\label{f1}
\end{equation}
We treat the second term in \fer{1.25} in a similar fashion. 
\begin{equation}
\begin{split}
\int_{{\mathbb R}}S_2(x)\d x = &\int_{{\mathbb R}} (1+|x|)^{-\frac{1}{2}-\eta}\scalprod{\psi_{
\theta}}{(L_{0\theta}^*-\bar
z)^{-\frac{1}{2}}(L_{0\theta}-z)^{-\frac{1}{2}}\psi_{
\theta}}\d x\\
=&\sum_{n=0}^\infty
\int_{{\mathbb R}}\big[\int_{{\mathbb R}}\frac{(1+|x|)^{-1/2-\eta}}{\sqrt{(l-x)^2+(\theta
n+1)^2}}\d x\big] \d \nu_n(l),
\label{hs1}
\end{split}
\end{equation}
where $\d \nu_n$ is the spectral measure of $L_\r$ associated to
vector $P(N=n)\psi_{\theta}$. The integral over $x$ is bounded above by
$$
\int_{{\mathbb R}}\frac{(1+|x|)^{-1/2-\eta}}{\sqrt{(l-x)^2+1}}\d x \leq \int_{{\mathbb R}}\big\{ (1+|x|)^{-1-2\eta} + \frac{1}{(l-x)^2+1}\big\}\d x\leq C_\eta +\pi,
$$
uniformly in $l\in{\mathbb R}$. It follows from the last estimate and \fer{hs1} that 
\begin{equation}
\int_{{\mathbb R}}S_2(x)\d x \leq (C_\eta+\pi) \|\psi_\theta\|^2<\infty.
\label{f2}
\end{equation}
The bounds \fer{f1} and \fer{f2} finish the proof that the integral on the right side of \fer{mm6} converges.

In order to complete the proof of Proposition \ref{prop5} (and hence that of Theorem \ref{thm2}), we need to show that the integral on the right side of \fer{mm6} is analytic in $s$, for $|s|<\frac{1}{2}+\epsilon$. To do so, let $\nu>0$ and set
\begin{equation}
\begin{split}
F_\nu(s)=\frac{-1}{2\pi\i}\int_{-\nu-\i}^{\nu-\i}\e^{\i tz} \scalprod{\phi_{\overline\theta}}{(K_\theta^{(s)}-z)^{-1}\psi_\theta}\d z,
\end{split}
\end{equation}
which is analytic in $s$, for  $|s|<\frac{1}{2}+\epsilon$. Denote by $F(s)$ the right side of \fer{mm6}. We have
\begin{equation}
\big|F_\nu(s)-F(s)\big|=\frac{1}{2\pi}\left|\Big(\int_{-\infty-\i}^{-\nu-\i}+\int_{\nu-\i}^{\infty-\i}\Big)\e^{\i tz} 
\scalprod{\phi_{\overline\theta}}{(K_\theta^{(s)}-z)^{-1}\psi_\theta}
\d z\right|.
\label{2.35}
\end{equation}
The above analysis shows that the integrals converge uniformly in $s$ and hence \fer{2.35} converges to zero uniformly in $s$. Therefore, $F(s)$ is analytic. This completes the proof of Proposition \ref{prop5} and that of Theorem \ref{thm2}. \qed

\bigskip
\noindent
{\bf Acknowledgement.}  This work has been supported by an NSERC Discovery Grant (Natural Sciences and Engineering Research Council of Canada).

\end{document}